\documentclass[prb,superscriptaddress,twocolumn,amsmath,amssymb]{revtex4}
\usepackage{graphicx}
\usepackage{multirow}

\begin{document}

\title{X-ray Resonant Scattering Study of the Order Parameters in Multiferroic TbMnO$_3$}

\author{D. Mannix}
\affiliation{(XMaS CRG Beam line, European Synchrotron Radiation Facility, F-38043 Grenoble, France)}
\author{D.F. McMorrow}
\affiliation{(London Centre for Nanotechnology and Department of Physics and Astronomy, University College London, UK)}
\affiliation{(ISIS Facility, Rutherford Appleton Laboratory, Chilton, Didcot OX11 0QX, UK)}
\author{R.A. Ewings}
\affiliation{(Department of Physics, Clarendon Laboratory, University of Oxford, UK)}
\author{A.T. Boothroyd} 
\affiliation{(Department of Physics, Clarendon Laboratory, University of Oxford, UK)}
\author{D. Prabhakaran} 
\affiliation{(Department of Physics, Clarendon Laboratory, University of Oxford, UK)}
\author{Y. Joly}
\affiliation{(CNRS-Grenoble, Grenoble, France)}
\author{B. Janousova}
\affiliation{(European Synchrotron Radiation Facility, F-38043 Grenoble, France)}
\author{C. Mazzoli}
\affiliation{(European Synchrotron Radiation Facility, F-38043 Grenoble, France)}
\author{L. Paolasini} 
\affiliation{(European Synchrotron Radiation Facility, F-38043 Grenoble, France)}
\author{S.B. Wilkins}
\affiliation{(European Synchrotron Radiation Facility, F-38043 Grenoble, France)}

\begin{abstract}
We report on an extensive investigation of the multiferroic compound TbMnO$_3$ using 
x-ray scattering techniques. Non-resonant x-ray magnetic scattering (NRXMS) 
was used to characterise the domain population of 
the single crystal used in our experiments. This revealed that the dominant domain is overwhelmingly 
$A$-type. The temperature dependence of the intensity and wavevector associated with the incommensurate magnetic 
order was found to be in good agreement with neutron scattering data. X-ray resonant scattering experiments were 
performed in the vicinity of the Mn $K$ and Tb $L_3$ edges in the high-temperature collinear phase, the 
intermediate temperature cycloidal/ferroelectric phase, 
and the low-temperature phase. In the collinear phase, where according 
to neutron diffraction only the Mn sublattice is ordered, resonant $E1-E1$
satellites were found at  the Mn $K$ edge associated with $A$-type but also $F$-type peaks. Detailed measurements
of the azimuthal dependence of the $F$-type satellites (and their absence in the NRXMS experiments) 
leads us to conclude that they are most likely non-magnetic 
in origin. We suggest instead that they may be associated with an induced charge multipole. At the Tb $L_3$ edge
resonant $A$- and $F$-type satellites were observed in the collinear phase again associated with $E1-E1$ events. 
These we attribute to a polarisation of the 
Tb 5$d$ states by the ordering of the Mn sublattice. On cooling into the cycloidal/ferroelectric 
phase a new set of resonant 
satellites appear corresponding to $C$-type order. These appear at the Tb $L_3$ edge only. In addition to a dominant
$E1-E1$ component in the $\sigma-\pi^\prime$ channel, a weaker component is found in the pre-edge 
with $\sigma-\sigma^\prime$
polarization and displaced by $-7$ eV with respect to the $E1-E1$ component. Comprehensive 
calculations of the x-ray scattering cross-section were performed using the $FDMNES$ code. These calculations  
show that the 
unrotated $\sigma-\sigma^\prime$ 
component of the Tb $L_3$ $C$-type peaks appearing in the ferroelectric phase contains a contribution from 
a multipole that is odd
with respect to both space and time, known in various contexts as the anapole. 
Our experiments thus provide tentative evidence 
for the existence of a novel type of anapolar order parameter in the rare-earth manganite class of 
mulitferroic compounds. 
\end{abstract}

\maketitle

\subsection{Introduction}
Materials possessing more than one "ferroic" order (which includes magnetism, ferroelectricity, 
elastic distortions, etc.) are referred to as multiferrroic. 
Although multiferroics are rare $-$  magnetism requiring a partially filled 
electronic shell, while canonical ferroelectrics such as BaTiO$_3$ have 
an empty one $-$ they are of considerable interest both from 
a fundamental point of view, and from the possibilities 
that they offer in the field of spintronics. 
Recently the RMnO$_3$ and RMn$_2$O$_5$ (R=rare earth) series of compounds 
have attracted a considerable amount of attention, 
as it has been shown that members of these series 
are multiferroic and display novel effects \cite{kimura2003,fiebig2002a,lottermoser2004b,goto2004,chapon2004,hur2004b}. 
In the case of TbMnO$_3$, for example, a giant magnetoelectric 
effect has been reported allowing the electric 
polarization to be switched by applied magnetic fields \cite{kimura2003}. 
Correspondingly, it has been shown in HoMnO$_3$ that the magnetic response can be 
controlled by an applied electric field\cite{lottermoser2004b}. 
Of key importance in understanding the
functionality of these materials is to obtain a complete description of the
various order parameters.

To shed further light on the nature of the order parameters in TbMnO$_3$ we 
have employed x-ray scattering techniques, 
with an emphasis on exploiting the 
rich possibilities offered by x-ray resonant scattering (XRS).
In an XRS experiment, the photon 
energy is tuned close to an absorption edge and an excited intermediate state is 
created by transitions between core and valence shells. 
At resonance, the scattering amplitude of 
a single ion can be described by a tensor, whose properties  are related directly 
to specific terms in the multipolar expansions of the electric charge and magnetization\cite{carra1994}.
XRS is now well established 
as being capable of revealing 
a great diversity of ordering phenomena in solids.
These are usually probed via the
pure excitation channels $E1-E1$ and $E2-E2$, where
the two distinct channels arise from selection rules for the change in angular
momentum $\Delta L$ between the core and excited states $\Delta L(E1-E1)=\pm 1$ and $\Delta L(E2-E2)=\pm 2$, respectively.
For such events, the tensors are even parity and are ranked according to their behaviour under
time reversal as time-odd, for magnetic, and time-even for charge orderings.
If, however, the ions are located in crystallographic positions which lack a 
centre of inversion symmetry 
(odd parity),
hybridisation will occur between the valence orbitals of that atom and transitions
may then occur via mixed processes 
to the hybridised states, giving rise to parity odd $E1-E2$ events. Such events
open up new and interesting possibilities for studying ordering phenomena driven by parity odd multipoles. 
For example, magnetoelectric toroidal moments may develop which can be
visualised as arising from placing a series of magnetic dipole moments end 
to end so as to form a closed circular path\cite{dubovik1990}. Such moments are characterised 
as having odd symmetry with respect to parity and can have odd or even symmetry in time reversal.
These toroidal moments have historically been referred to as anapoles first considered in the context of
multipolar  expansions in nuclear physics\cite{zeldovich1957}.
Multipoles associated with $E1-E2$ processes 
have been shown to be of importance in non-reciprocal and natural 
circular dichroism\cite{goulon1998, goulon2000, carra2003}, and recent 
predictions having been made for XRS experiments\cite{marri2004,dimatteo2005,dimatteo2006,lovesey2005}.

\begin{figure}[t]
\centering
\includegraphics[width=0.45\textwidth,clip]{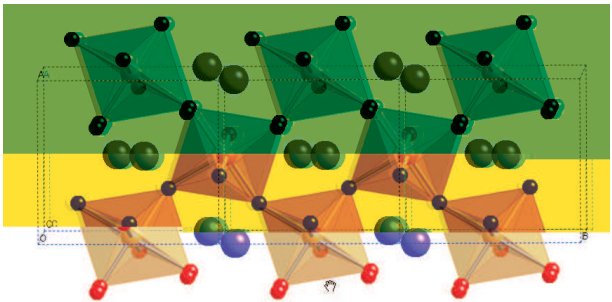}
\caption{(Color online) {\bf a} The crystal structure of TbMnO$_3$ in its high-temperature (Pbnm) phase
shown in projection down the ${\bf c}$ axis. 
}
\label{schematic}
\end{figure}

At room temperature TbMnO$_3$ adopts an orthorhombic structure, space group $Pbnm$, 
with lattice parameters a=5.316 \AA, b=5.831 \AA\ and c=7.375 \AA\ (Fig.\ \ref{schematic}). 
The Mn and Tb ions occupy the $4b$ (point symmetry $\overline{1}$, i.e. inversion symmetry)
and $4c$ (point symmetry $m$) sites, respectively.
Neutron diffraction experiments have shown 
that the Mn$^{3+}$ magnetic moments order below $T_{N1}$=41 K to form 
an incommensurate, longitudinally polarized structure 
(hereafter referred to as the collinear phase) described by a wave-vector 
$(0,\mathrm q_{Mn},0)$ with q$_{Mn}$=0.28-0.29 b$^*$ \cite{kenzelmann2005,kajimoto2004}.
In this collinear phase the Tb moments are thought to be disordered \cite{kenzelmann2005}.
Below $T_{N2}=28$~K the magnetic structure becomes 
non-collinear (hereafter referred to as the cycloidal phase) by developing 
a component of the magnetization on the Mn sublattice along the ${\bf c}$ axis, and at exactly the 
same temperature the material becomes ferroelectric \cite{kenzelmann2005}. 
According to modelling of the neutron diffraction, the Tb moments in this phase order at the same wavevector as
the Mn moments, but are transversely polarized along 
the ${\bf a}$ axis. On further cooling below $T_{N3}=7$~K, 
the Tb moments undergo a further transition and develop a component of the magnetization at 
a distinct wave-vector of $(0,\mathrm q_{Tb},0)$ 
with q$_{Tb}$=0.42 b$^*$, again thought to be transversely polarized along the ${\bf a}$ axis.  
Kenzelmann {\it et al.} \cite{kenzelmann2005} 
have argued that the transition to a ferroelectric
state results from a loss of a centre of inversion symmetry when the 
magnetic structure changes from being collinear to non-collinear at $T_{N2}=28$~K. 
This idea has generated considerable interest, and ferroelectric transitions driven 
by the formation of non-collinear magnetic structures 
have now been reported in a number of systems\cite{yamasaki2006,lawes2005}.
Several different theoretical approaches have been developed to explain the magneto-electric properties of
this class of multiferroic\cite{harris2006,katsura2005}, which suggest that the macroscopic electric polarization
${\bf P}$ is related to the magnetic moment ${\bf S}_i$ by
\begin{equation*}
\mathbf P= a\,{\mathbf e}_{ij}\times(\mathbf S_i\times \mathbf S_j)
\end{equation*}
where $e_{ij}$ is a vector connecting the spins on the sites $i$ and $j$, and $a$ is a constant.

It should be noted that a certain degree of 
inconsistency exists in the neutron diffraction 
literature concerning the magnetic structure adopted by TbMnO$_3$. 
In TbMnO$_3$ the magnetic reflections may be classified into four distinct types, which have been shown to 
have distinct temperature dependences  \cite{wollan,bertaut,kenzelmann2005,kajimoto2004}. 
The $F$-type structure gives rise 
to magnetic reflections classified by $h+k$=even, $l$=even;
the $G$-type reflections with $h+k$=odd, $l$=odd; the $C$-type $h+k$=odd and $l$=even;
and the $A$-type structure gives reflections corresponding to $h+k$=even, $l$=odd.  
In the original
investigation by Quezel {\it et al.}\cite{quezel1977} 
strong $A$-type and weak $G$-type magnetic peaks were reported, with the 
systematic absence of 
$C$-type and $F$-type reflections. 
Blasco {\it et al.}\cite{blasco2000} obtained similar results from powder neutron diffraction. In their neutron
work on a large single crystal of TbMnO$_{3}$, Kajimoto {\it et al.}\cite{kajimoto2004} 
found evidence for all $A$-, $F$-, $C$- and $G$-type
structures. The $A$ and $G$-type structures have their moments unambiguously along the ${\bf b}$ axis,
but attempts to determine the moment directions
for the $F$ and $C$-type structures could only constrain the direction to be either the ${\bf c}$ or ${\bf a}$ axis. 
The more recent neutron investigation by Kenzelmann {\it et al.}\cite{kenzelmann2005} reported
only the $A$-type reflections with moments along the ${\bf b}$ axis in the collinear phase, in agreement with 
Kajimoto {\it et al}. 
A plausible explanation for the observation
of the different magnetic structures in these experiments is that they depend on details of sample growth and 
preparation. 
All neutron investigations, however,  agree on one thing: 
the $A$-type structure is always the dominant phase. It is also worth noting that
the magnetic order is accompanied by weak charge satellites at 2q$_{Mn}$ which
have been observed with neutron and non-resonant x-ray scattering \cite{arima2005,aliouane2006}. 

Our experiments on TbMnO$_3$ were undertaken in the spirit of exploring 
whether the well documented characteristics
of XRS could be exploited to reveal
information complementary to that provided by neutron diffraction. In particular, 
the element and electron shell specificity of x-ray resonant magnetic scattering (XRMS) is often used to isolate the 
contribution from individual components in systems with more than one type of magnetic 
species, such as TbMnO$_3$. A second major objective of our study was to investigate whether 
the loss of inversion symmetry that must accompany the transition to the ferroelectric state
in TbMnO$_3$ allows new terms 
in the XRS cross-section to become visible,
such as $E1-E2$ terms associated with the ordering of parity odd 
multipoles\cite{marri2004,dimatteo2005,dimatteo2006,lovesey2005}. 

This paper is organised as follows. In the next section, \ref{sec:exp}, we outline the experimental details,
including a description of the scattering geometries (crystal orientation, 
x-ray polarisation, definition of azimuth, {\it etc.}) employed in our experiments. In
Sec.\ \ref{sec:results} we present the experimental results, 
which for clarity of exposition is further sub-divided into sub-sections. Our attempts
to characterise the magnetic domain structure of our sample using non-resonant x-ray magnetic scattering
(NRXMS) are described in Sec.\ \ref{sec:nrxms}. The 
results of the XRS investigations in the collinear phase
at the Mn $K$-edge are described in Sec.\ \ref{sec:co_mn}, while in Sec.\ \ref{sec:co_tb} we describe
the outcome of XRS experiments in the 
collinear phase at the Tb $L_{3}$ edge. XRS investigations
undertaken in the cycloid phase, again at the Mn $K$ and  Tb $L_{3}$ edges, are described in 
Sec.\ \ref{sec:cy_mn} and Sec.\ \ref{sec:cy_tb}, respectively. Results taken in the 
the low temperature regime, $T<8K$, at the Tb $L_3$ edge are
presented in Sec.\ \ref{sec:lt_tb}. The investigation of the 2q charge satellites is presented 
in Sec.\ \ref{sec:2q}. Our  results are analysed and
discussed in Sec.\ \ref{sec:ana}, in the light of calculations of the electronic structure 
using the $FDMNES$ code\cite{joly2001}. In the final section, \ref{sec:sum}, 
we summarise our work.   

\begin{figure}[t]
\centering
\includegraphics[width=0.45\textwidth,,bb=197 378 476 667,,clip]{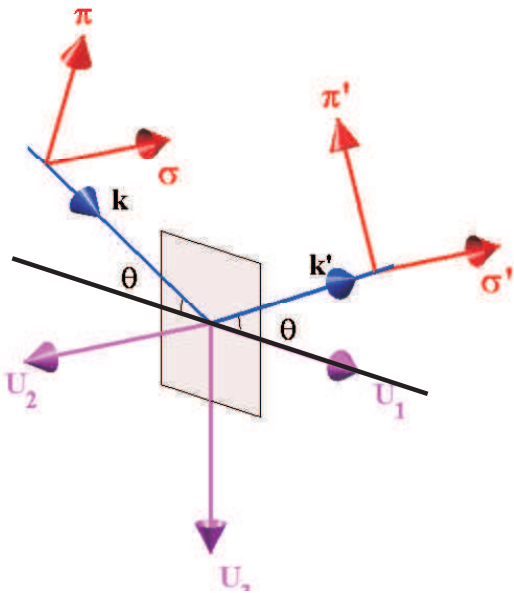}
\caption{ (Color online) Definitions of the nomenclature used to label the polarization state of the photon beam and the coordinate 
system used to resolve the components of the magnetic moments relative to the incident and scattered beams.
}
\label{u1u2u3}
\end{figure}

\subsection{Experimental Details}
\label{sec:exp}
Single crystals of TbMnO$_{3}$ were prepared in Oxford by the floating zone method in an image furnace. 
The crystals were cut 
and polished to produce samples with either 
the  ${\bf b}$ or ${\bf a}$ axis as the surface normal. 
The samples were investigated by SQUID magnetometery and were found to display the same 
behaviour as previously reported \cite{kimura2003b}. The lattice parameters refined in our
synchrotron experiments were in very good agreement with with those deduced
from neutron scattering results and the our crystal mosaic was found to be $\approx{0.03}$ degrees.
These observations indicate that the samples used in our studies were of execellent quality
with comparable bulk magnetic properties to other samples described in literature.   
      
The scattering experiments were undertaken at the 
XMaS and ID20 magnetic scattering beamlines at the ESRF, Grenoble, France. 
XMaS views radiation from a bending magnet and employs   
a vertical scattering geometry ($\sigma$-polarised incident photons). The sample used on XMaS was cut and polished  
with a $\mathbf{b}$ axis as surface normal (see Fig.\ \ref{u1u2u3} for the definition of 
photon polarizations and coordinate system used in our experiment). 
The experiments on the undulator beamline ID20 were performed also mainly 
with the same vertical geometry as on XMaS. Some additional 
experiments on ID20 were carried out in a normal beam geometry,
with the scattering plane approximately horizontal, corresponding to $\pi$-polarised incident photons. 
XRS measurements were undertaken around the Mn $K$ (6.552 keV) 
and Tb $L_{3}$ (7.515 keV) edges, using a Cu(220) or Au(222) polarisation analyser, respectively. 
The wave-vector
resolution was of the same order of magnitude in all our x-ray experiments, typically 
$\approx$  1$\times$10$^{-4}$ reciprocal lattice units (r.l.u.).
The photon flux on ID20 is approximately 1$\times$10$^{13}$ per second 
in a focused spot size of about 0.3$\times$0.4mm$^2$ (vertical $\times$ horizontal); 
on XMaS the corresponding numbers are 5$\times$10$^{11}$ per second in 
0.8$\times$0.5mm$^2$ at x-ray energy of 8keV and 200 mA current in the ESRF storage ring.
The intense photon flux on ID20 carries with it the potential to 
heat the sample. Checks were made for such an effect, and when present (typically for sample temperatures
below 10 K), the incident beam was attenuated by up to an order of magnitude.
A key aspect of our experiments has been the utilisation of the high flux from the ID20 undulator source,
which enabled us to perform non-resonant x-ray magnetic scattering (NRXMS) to characterise the
magnetic domains present in the near surface volume probed by the x-rays. These experiments were undertaken
with incident energy of 7.470 keV using a Au(222)polarisation analyser.

In the following, we use the  conventional notation to describe the polarisation conditions of
the incident and scattered photon beams, $\sigma$-$\sigma^\prime$ and $\pi$-$\pi^\prime$ for incident
polarisation scattered without rotation and $\sigma$-$\pi^\prime$ and $\pi$-$\sigma^\prime$ for
photons scattered with a 90$^\circ$-rotation of their electric vector.

\begin{figure}[t]
\centering
\includegraphics[width=0.40\textwidth,clip]{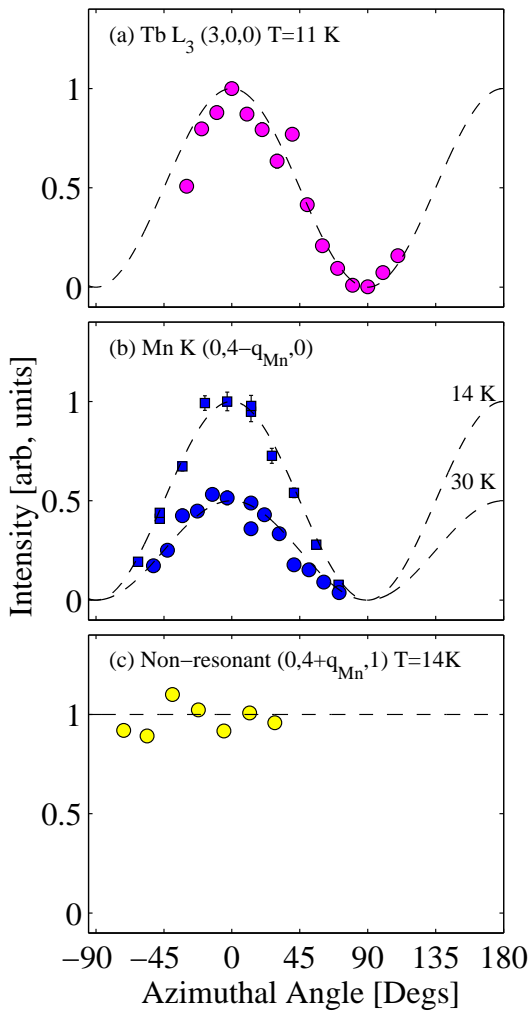}
\caption{(Color online) 
Summary of azimuthal scans of various diffraction peaks
recorded at different temperatures and photon energies.
The data were recorded in the $\sigma-\pi^\prime$
channel and have been normalised to unity. 
The lines are guides to the eye.
}
\label{az_s1}
\end{figure}

In our experiments extensive use was made of scans of the azimuthal angle $\psi$, ${\it i.e.}$ 
rotation around the scattering vector \cite{finkelstein1992}. 
In recent years the
azimuthal dependence of XRS has been exploited increasingly in studies of systems displaying
magnetic and multipolar order. The reason is that the variation
of the resonant intensity in a scan of the azimuthal angle reflects
directly the symmetry of the intermediate states (orbitals) through which the
resonance proceeds. By way of example, we show in Fig.\ \ref{az_s1}(a) the 
azimuthal dependence of the space-group forbidden (3,0,0) reflection recorded
at the Tb $L_3$ edge on XMaS. The intensity displays a $\cos^2\psi$ dependence in agreement
with calculations of $E1-E1$ Templeton scattering\cite{templeton1982, murakami1998} for the $Pbnm$ space group.
In either the experiments on samples with ${\bf b}$ or ${\bf a}$
faces as the surface normal direction, 
the crystal was oriented with the ${\bf c}$ axis in the scattering plane, which defined the $\psi$=0 of the
azimuthal scans. Due to the construction of the diffractometers used in our experiments, 
azimuthal scans could only be performed in the vertical scattering geometry, and 
then mainly for reflections parallel to the ${\bf b}$ axis. 

\subsection{Experimental Results}
\label{sec:results}

\subsubsection{Magnetic Structure Investigation with NRXMS}
\label{sec:nrxms}

In order to ascertain which magnetic structures were present in our sample, and in particular to
understand any effects that may arise in the near surface region probed
in our x-ray measurements, we initially investigated the non-resonant x-ray 
magnetic scattering (NRXMS).
Time constraints, and the weakness of the signal, prevented us from making exhaustive 
studies in the NRXMS regime. 
Nevertheless, these investigations turned out to be helpful in the interpretation of our XRS results. 

The NRXMS experiments were undertaken at a temperature of $T$=14~K where all the 
possible $A$-, $F$-, $C$- and $G$-type reflections were found to be present in the study of 
Kajimoto {\it et al.}\cite{kajimoto2005}. The crystal was orientated initially at an azimuth of $\psi=0$, and a 
photon energy of 7.470 keV was selected, below the Tb $L_3$ edge. 
The scattered intensity calculated for NRXMS can be written as \cite{blume1988}
\begin{align}
I^{\sigma-\sigma^\prime} &= V_{d} (S_{2}\sin2\theta)^2 \notag \\ 
I^{\sigma-\pi^\prime} &= V_{d} ((-2\sin^2\theta)[(\cos\theta)(L_{1} \notag \\
                      & + S_{1})-S_{3}\sin\theta])^2  
\label{eq:nrxms}
\end{align}
where $L_{1}$, $S_{1}$, $S_{2}$ and $S_{3}$ are the components of orbital and 
spin moments along orthogonal axes $U_1$, $U_2$ 
and $U_3$ (Fig.\ \ref{u1u2u3}), $\theta$ is the scattering Bragg angle,
and V$_{d}$ represents the domain volumes of the $d=A$, $F$, $C$,  or $G$-type 
structures within the illuminated sample volume. Upon azimuthal rotation, consideration
should be given in  Eq.\ \ref{eq:nrxms} for the change in the magnetisation components along the $U_1$, $U_2$ and $U_3$ axes,
so that for example, $S2(\psi)=S2(\psi=0)\cos\psi - S1(\psi=0)\sin\psi$.  

\begin{figure}[t]
\centering
\includegraphics[width=0.35\textwidth,clip]{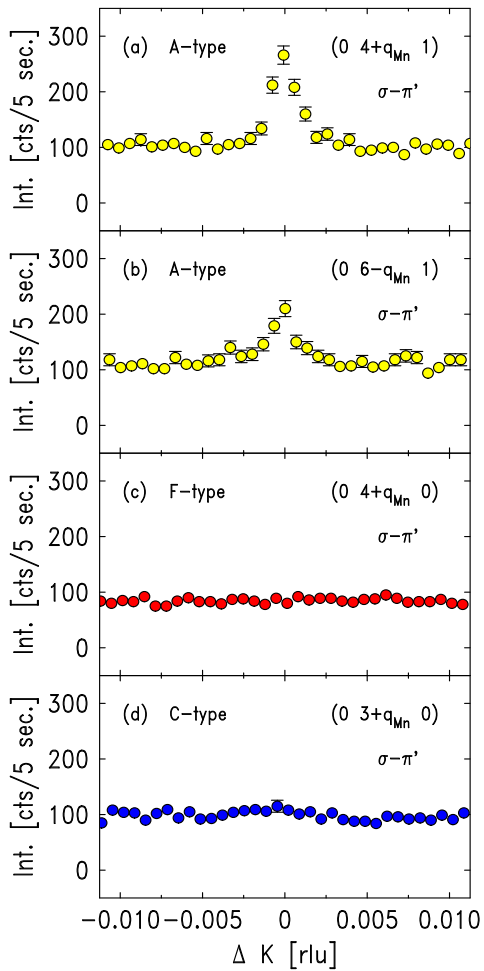}
\caption{ (Color online) 
Summary of the non-resonant x-ray magnetic scattering
observed on ID20 in scans of wavevector transfer parallel to ${\bf b}^*$ 
through various positions in reciprocal space associated with  
the $A$-, $F$-, or $C$-type 
magnetic structures. The data were 
recorded in the ferroelectric/cycloidal phase at  
$T=14 K$, at an azimuth of $\psi=0$ and in the $\sigma-\pi^\prime$ channel.
}
\label{nrsp}
\end{figure}

\begin{figure}[t]
\centering
\includegraphics[width=0.35\textwidth,clip]{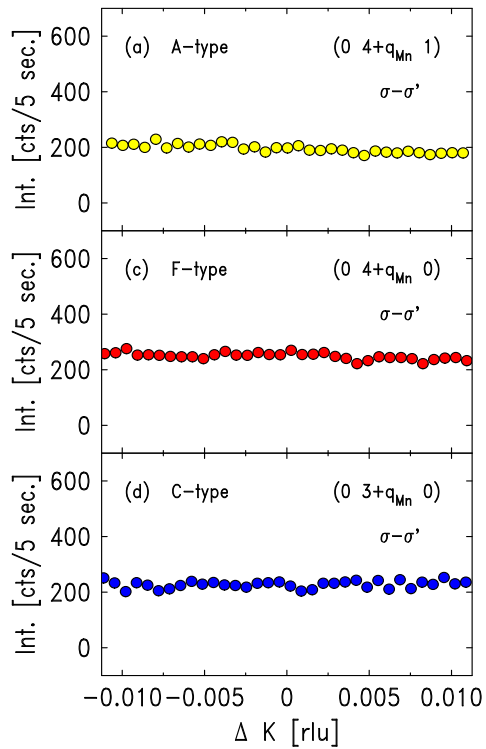}
\caption{ (Color online) 
Summary of the non-resonant x-ray magnetic scattering
observed on ID20 in scans of wavevector transfer parallel to ${\bf b}^*$ 
through various positions in reciprocal space associated with  
the $A$-, $F$-, or $C$-type 
magnetic structures. The data were 
recorded in the ferroelectric/cycloidal phase  
at $T=14 K$, at an azimuth of $\psi=0$ and in the $\sigma-\sigma^\prime$ channel.
}
\label{nrss}
\end{figure}

The results of our NRXMS investigations are summarised in Fig.\ \ref{nrsp}
and Fig.\ \ref{nrss} for the $\sigma$-$\pi^\prime$ and $\sigma$-$\sigma^\prime$ 
polarised intensities, respectively. From Eq.\ \ref{eq:nrxms},
it can be seen that the $\sigma$-$\pi^\prime$ channel is sensitive to magnetic 
moments in the scattering plane along  $U_1$ and $U_3$,
and the $\sigma$-$\sigma^\prime$ channel is sensitive to magnetic moments 
directed perpendicular to the scattering plane along $U_2$.
In Fig.\ \ref{nrsp}, significant NRXMS intensity is evident for wavevectors 
corresponding to the $A$-type magnetic structure, $(0,4\pm\mathrm q,1)$
and $(0,6\pm\mathrm q,1)$. 
(The slightly weaker intensities found at $(0,6\pm\mathrm q,1)$ compared to $(0,4\pm\mathrm q,1)$ is

probably due to the decreasing magnetic form-factor rather than the geometric 
terms in Eq.\ \ref{eq:nrxms}.) Scans, taken at
wavevectors corresponding to the $F$-type and $C$-type structures, at 
$(0,4\pm\mathrm q,0)$ and $(0,3\pm\mathrm q,0)$ are shown in Fig.\ \ref{nrsp}(c) and (d),
where no peaks are visible above background. Turning to the 
$\sigma$-$\sigma^\prime$ channel, summarised in Fig.\ \ref{nrss}, we found no evidence for NRXMS
for the wavevectors investigated, indicating that the magnetic moments in this phase 
are confined to the ${\bf b}-{\bf c}$ plane. 

Thus the fact that we observe satellite reflections in our NRXMS experiments 
at positions $(h,k\pm\mathrm q,l)$, with $h+k$ even, $l$ even,  appears to indicate that 
our sample is predominantly $A$-type. While the fact that we do not 
observe any peaks in the unrotated $\sigma-\sigma^\prime$ channel leads us to conclude 
that in our sample, and at a temperature well inside  the 
cycloidal phase, the magnetic moments are in the ${\bf b}-{\bf c}$ plane.
In this way our sample appears to share a similar domain configuration to the one studied 
by Kenzelmann {\it et al.} who found that in this phase the moments form a cycloid in the 
${\bf b}-{\bf c}$ plane. Further information on the orientation of magnetic moments 
in our sample was provided by studying the azimuthal dependence of the $(0,4\pm\mathrm q,1)$ satellite, 
shown in Fig.\ \ref{az_s1}(c). The intensity depends only weakly on azimuth, which from Eq.\ \ref{eq:nrxms},
implies that the moments have their strongest component along the ${\bf b}$ direction. It should be noted
that the weak azimuth dependence observed could also arise from 2 magnetic domains oriented along both the
a and c axes in our scattering geometry. However, the absence of magnetic scattering in $\sigma-\sigma^\prime$
channel is not consistent with such a scenario. 

\begin{figure}[t]
\centering
\includegraphics[width=0.40\textwidth,clip]{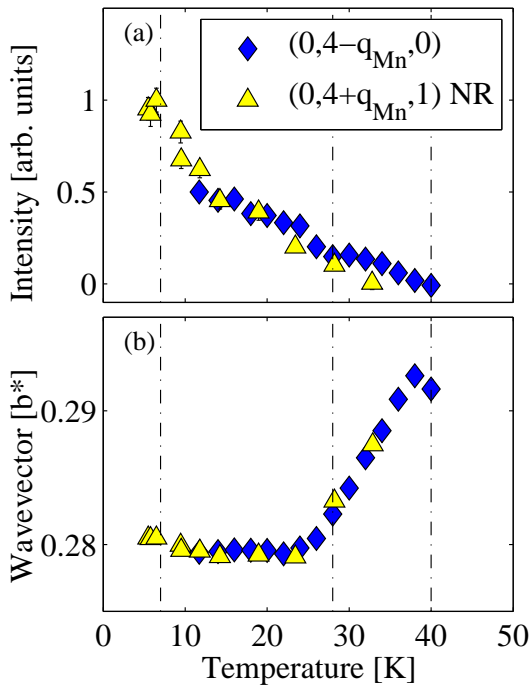}
\caption{ (Color online) 
Temperature dependence of the intensity and modulation wave-vector
of various satellites. In (a) and (b)
we compare Mn $K$ edge and non-resonant data, all recorded in 
$\sigma-\pi^\prime$. For ease of comparison, the intensities have been normalised to unity 
at low temperature. The data were taken at ID20.
}
\label{t_1}
\end{figure}

The temperature dependence of the intensity of NRXMS observed at 
$(0,4\pm\mathrm q,1)$ is shown in Fig.\ \ref{t_1}(a), where it appears to
agree with that deduced from neutron scattering experiments, increasing steadily below $T_{N1}$=41~K.
We also find good agreement with neutron scattering 
for the thermal evolution of the magnetic wave-vector 
as shown in Fig.\ \ref{t_1}(b),
starting at a value of about $0.29$ at T$_{N1}$, decreasing in value 
as the sample is cooled through the collinear phase, before 
appearing to lock into a value of about $0.28$ below T$_{N2}$ in the ferroelectric phase. 
However, in agreement with the neutron data, it is clear that this is not a genuine 
lock-in transition, as below $T_{N2}$ the value of the wavevector 
increases slightly down to base temperature.

The conclusion to be drawn from our NRXMS investigation is 
that the dominant magnetic structure in our sample appears to be $A$-type, with the magnetic
moment oriented mainly along the ${\bf b}$ axis. No evidence is  found 
for the $F$-type or $C$-type structures, presumably because either they are not present, or
their domain volumes are too small to allow them to be observed.

\begin{figure}[t]
\includegraphics[width=0.40\textwidth,clip]{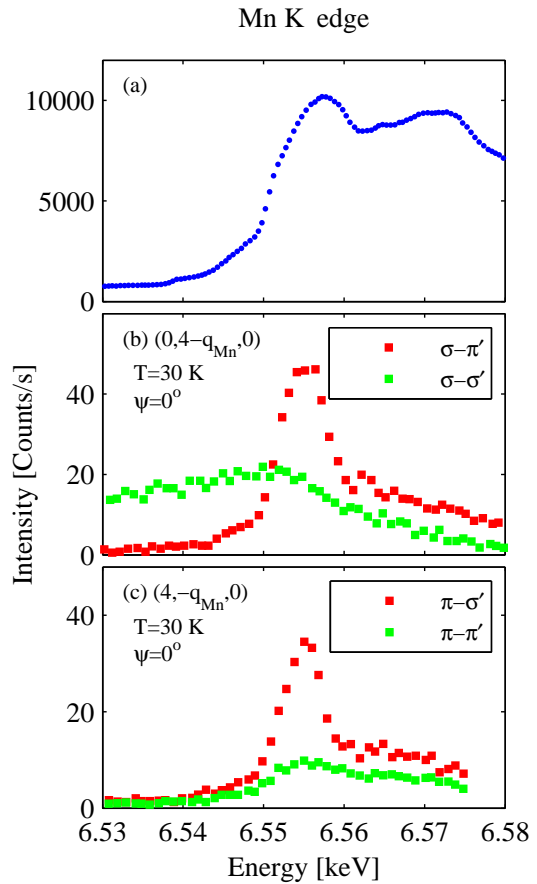}
\caption{ (Color online) (a) Fluorescence spectrum in the vicinity of the Mn $K$ edge. (b)-(c) 
Representative energy dependence of the x-ray resonant scattering 
near the Mn $K$  edge at various satellite positions, and for 
various polarization geometries. The data were taken at ID20.
}
\label{es_mn1}
\end{figure}

\begin{figure}[t]
\includegraphics[width=0.45\textwidth,bb=4 335 306 646,clip]{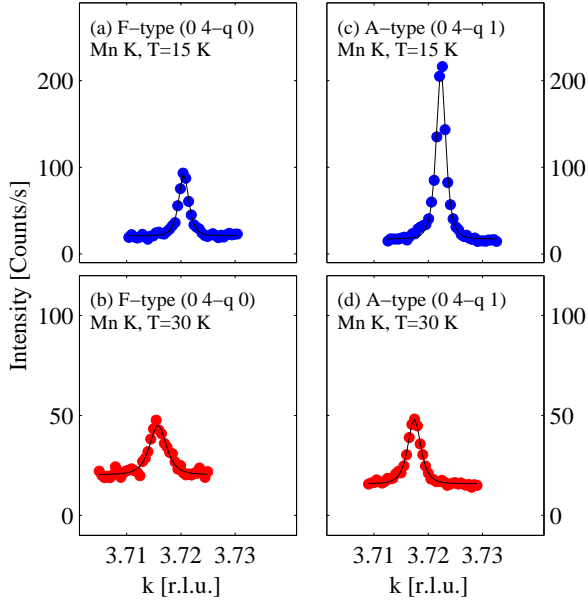}
\caption{ (Color online) 
Representative scans of the wavevector transfer with the
photon energy tuned to the peak of the resonance at the Mn K edge. The scans shown are through the 
positions of $F$-type ((a) and (b)) and $A$-type ((c) and (d)) reflections, and were recorded in the 
$\sigma-\pi^\prime$ channel at ID20.
}
\label{kscans_mn}
\end{figure}

\subsubsection{Collinear phase: Mn K-edge}
\label{sec:co_mn}

Turning to our XRS studies, we first investigated the response with incident energy tuned close to the maximum of the Mn 
fluorescence spectrum (Fig.\ \ref{es_mn1}(a)), and with the sample cooled into the collinear
phase ($T_{N2}\leq T \leq T_{N1}$). 
Under these conditions, scans in reciprocal space were performed to search for the existence of satellite peaks.
Well defined diffraction satellites were found (Fig.\ \ref{kscans_mn}(c) and (d)) 
at nominal $A$-type magnetic wave-vectors such as $(0,k\pm\mathrm q,1)$,
with $\mathrm q\approx 0.285$ b$^*$, close to the value of q$_{Mn}$ deduced from neutron diffraction experiments. 
An additional set of  satellites were discovered at 
$F$-type positions, $(0,k\pm\mathrm q,0)$  $k$ even (Fig.\ \ref{kscans_mn}(a) and (b)). 
Both sets of satellites were found in the rotated $\sigma-\pi^\prime$ channel only.
Extensive searches at $C$-type positions such as $(0,3\pm\mathrm q,0)$, 
failed to find any evidence for XRS at the Mn $K$ edge in this phase. 
The observation 
of both $A$- and $F$-type reflections at the Mn K edge is difficult to reconcile with the 
NRXMS presented in Sec.\ \ref{sec:nrxms} where peaks attributable to the $A$-type structure only were
present. Moreover, the data in Fig.\ \ref{kscans_mn} also show 
similar intensity at the $A$- and $F$-type reflections in the collinear phase. 
This would appear to undermine any claim that the latter is magnetic scattering from a minority 
$F$-type phase: neutron scattering studies always show that $A$-type is  dominant.
We comment on possible explanations for the discrepancy between the NRXMS and XRS later. 

In Fig.\ \ref{es_mn1}(b) we show the energy dependence of the $(0,4-\mathrm q,0)$ at $30$ K, for incident $\sigma$ 
and the  $(4,-\mathrm q,0)$ for $\pi$-polarised photons (Fig.\ \ref{es_mn1}(c)),
as the photon energy is tuned through the Mn $K$ edge.
The resonant response is seen to exhibit a maximum at an energy of $6.557$ keV, slightly above the position of the first 
derivative in the absorption at
$6.552$ keV. The energy of the resonance 
allows us 
to identify unambiguously  the maximum with $E1-E1$ dipole transitions, which 
at the Mn K edge probe the $4p$ states. 
(A weak, pre-edge feature at $6.540$ keV is also evident.) The scattering was found to be mostly in 
the rotated channel. The scattering in the unrotated channel is more difficult to interpret with any 
certainty, due to changes in the diffuse charge scattering and the onset of fluorescence at the edge.

To further address the nature of the order revealed by the $F$-type peaks, the azimuthal 
dependence of the scattering of the $(0,4-\mathrm q,0)$ at the Mn $K$ edge was determined with the results 
shown in Fig.\ \ref{az_s1}(b) for the $\sigma$-$\pi^\prime$ channel. The scattering is seen 
to display the same  $\cos^2\psi$ dependence as the (3,0,0) space group forbidden reflection (Fig.\ \ref{az_s1}(a)).
The $E1-E1$ XRMS amplitude  can be expressed as \cite{hill1996} 
\begin{align}
\label{eq:E1res}
A_{res}^{Mag} & = 
\left(
\begin{smallmatrix}
\sigma-\sigma^\prime & \pi-\sigma^\prime\\
\sigma-\pi^\prime & \pi-\pi^\prime
\end{smallmatrix}
\right) \notag \\
&=
\left( 
\begin{smallmatrix} 
0 &  \hat{\mathrm{z}}_1\cos\theta + \hat{\mathrm{z}}_3\sin\theta \\
\hat{\mathrm{z}}_3\sin\theta-\hat{\mathrm{z}}_1\cos\theta & -\hat{\mathrm{z}}_2\sin2\theta 
\end{smallmatrix}
\right)
\end{align}
Here the amplitude is written  as a matrix in
a basis of the linear components
of the polarization perpendicular
($\sigma$) and parallel ($\pi$) to the scattering plane, for the
incident (unprimed) and scattered (primed) beams and
$\hat{\mathbf{z}}$ is a unit vector
parallel to the magnetic moment, 
with Cartesian components defined with respect to the orthogonal axes 
$U_1$, $U_2$ and $U_3$ (Fig.\ \ref{u1u2u3}). 
The above equation establishes
that $E1-E1$ XRMS is forbidden in the $\sigma-\sigma^\prime$ channel, and 
is sensitive to the magnetic moments in the
scattering plane for the $\sigma$-$\pi^\prime$ channel. Thus the two-fold azimuthal 
symmetry we observe at $(0,4-\mathrm q,0)$ (Fig.\ \ref{az_s1}(b)), with the 
maximum at {$\psi=0$}, could be consistent with magnetic scattering from an $F$-type structure for which the magnetic 
moments would have to be orientated along the ${\bf c}$-axis. 
This moment direction is in agreement with that deduced 
from neutron scattering experiments \cite{kajimoto2004}. However, when we compare our expectations for the NRXMS 
response that would arise from such an $F$-type
structure, with moments orientated along the ${\bf c}$ axis, we fail to find consistency with our own NRXMS results for 
the absence of signal at $(0,4+\mathrm q,0)$.
From Eq.\ \ref{eq:E1res}, using the Bragg angles 
at the Mn K-edge $\theta(0,4-\mathrm q,0)$=40 and $(0,4-\mathrm q,1)$=29.7,
the measured XRMS intensity ratio I($(0,4+\mathrm q,1)$)/I($(0,4+\mathrm q,0)$) can be used to 
obtain the expected volume ratio V$_{A}$/V$_{F}$ for the F-type and A-type structures.
This leads to the ratio V$_{A}$/V$_{F}$=3. We can then plug this ratio 
into Eq.\ \ref{eq:nrxms} for the non-resonant 
scattering described in Sec.\ \ref{sec:nrxms}, using the NRXMS Bragg angles $(0,4-\mathrm q,0)$=34.8 
and $(0,4-\mathrm q,1)$=23.8 
and calculate the NRXMS intensity we would expect at $(0,4+\mathrm q,0)$. 
However, from this we find that the NRXMS should be significantly larger than that observed at $(0,4+\mathrm q,1)$,
not zero as we have found. 

\begin{figure}[t]
\includegraphics[width=0.45\textwidth,bb=4 335 306 646,clip]{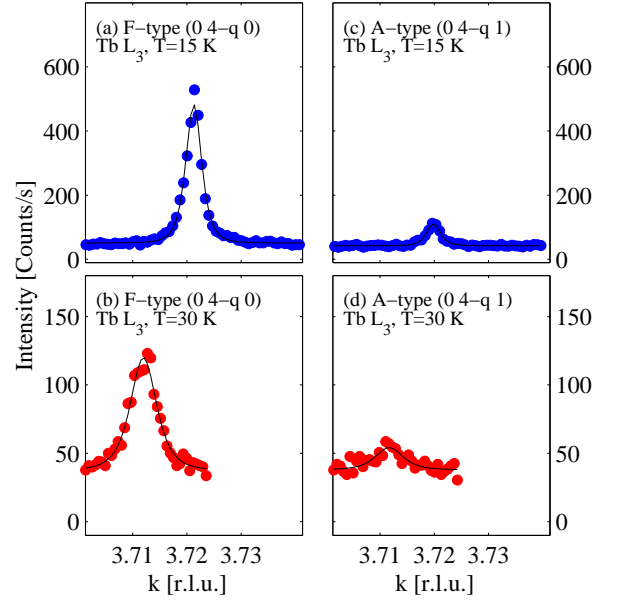}
\caption{ (Color online) 
Representative scans of the wavevector transfer with the
photon energy tuned to the peak of the resonance at the Tb $L_3$ edge. The scans shown are through the 
positions of $F$-type ((a) and (b)) and $A$-type ((c) and (d)) reflections, and were recorded in the 
$\sigma-\pi^\prime$ channel at ID20.
}
\label{kscans_tb}
\end{figure}

\begin{figure}[t]
\centering
\includegraphics[width=0.40\textwidth,clip]{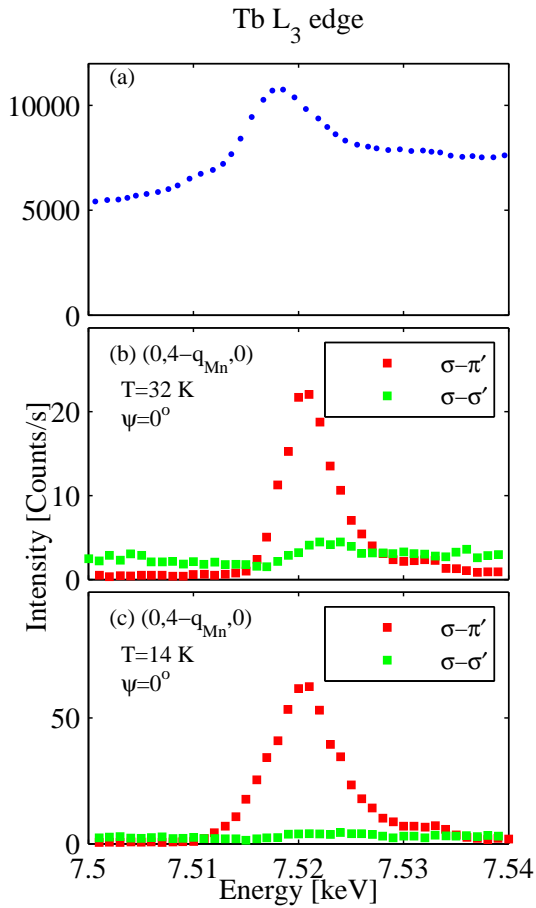}
\caption{(Color online) (a) Fluorescence spectrum in the vicinity of the Tb $L_3$ edges. 
(b)-(c) Representative energy dependence of the x-ray resonant scattering 
near the Tb $L_3$ edges at various satellite positions, and for 
various polarization geometries. The data were taken on XMaS.
}
\label{es_tb1}
\end{figure}

Further doubt that the XRS at $(0,4\pm\mathrm q,0)$ is magnetic in origin comes from its temperature dependence. 
The thermal evolution of the $(0,4-\mathrm q,0)$ $\sigma$-$\pi^\prime$ 
intensities are shown in Fig.\ \ref{t_1}(a) and appear to follow
closely the $A$-type $(0,4-\mathrm q,1)$ magnetic intensities that we 
measured using NRXMS. The close agreement in the temperature dependence 
is in contrast to that revealed in neutron diffraction experiments \cite{kajimoto2004}. In these measurements 
the temperature dependence for the two structures is very different: the F-type intensities increase at a slow 
rate below T$_{N1}$=41 K and then increase dramatically below $T_{N3}=7$~K due to the apparent stabilisation of the 
$F$-, $G$- and $C$-type structures by the onset of Tb magnetic order.

Also compared in Fig.\ \ref{t_1}(b) is the temperature dependence of the modulation wavevector at 
$(0,4-\mathrm q,0)$ with the non-resonant scattering at $(0,4-\mathrm q,1)$. 
Of significance, especially in the light of evidence that the XRS observed at 
$(0,4-\mathrm q,0)$ may not be magnetic in origin, is the similar thermal evolution of wave-vector to 
that found for $A$-type magnetic scattering we measured at $(0,4-\mathrm q,1)$ with NRXMS. This effect 
may suggest that the underlying order parameter is driven by the onset of magnetism of the Mn ions, 
suggesting an interesting coupling of magnetic and charge order parameters in this material.

In summary, in the collinear phase at the Mn $K$ edge we find $A$-type reflections 
at $(0,4-\mathrm q,1)$, which was the dominant peak in the NRXMS study.
Its dependence on energy, polarization, temperature, {\it etc.}, have the hallmarks of 
an XRMS process, where the ordering of the local Mn 3$d$ moments leads to a polarization 
of the $4p$ band. The peak at the $F$-type position $(0,4\pm\mathrm q,0)$ is less easy to interpret.
If it is attributed to also being an XRMS peak, then we can infer that for the $F$-type structure
the Mn moments are polarized along  the {\bf c} axis. However, the characteristics of 
the $(0,4\pm\mathrm q,0)$ $-$ not least of which is its absence in our NRXMS study $-$ forces us to
consider other possible origins for this peak. If this order parameter is not magnetic order, and given that it can be 
associated with $E1-E1$ events, we may suppose that the XRS arises from $E1-E1$ time-even events 
associated with a charge multipole. Moreover, given that 
its wavevector follows so closely with the magnetic, this order parameter would have to be coupled 
linearly to the magnetism of the Mn sublattice. 

\subsubsection{Collinear phase: Tb L$_{3}$ edge}
\label{sec:co_tb}

Experiments were also performed in the collinear phase for photon energies in the vicinity of
the Tb $L_3$ edge.
Resonant satellites were found at positions corresponding to both 
the nominal Mn $A$-type magnetic 
satellites at $(0,4\pm\mathrm q,1)$ and the seemingly anomalous peaks at $(0,4\pm\mathrm q,0)$. 
Representative scans of the wavevector transfer parallel to ${\bf b}^*$ are shown in 
Fig.\ \ref{kscans_tb}, where sharp diffraction peaks are evident reflecting long-range nature of the
ordering probed. Polarization analysis of these showed  
that the scattering was predominately in the rotated $\sigma$-$\pi^\prime$ channels.
The observation of these satellites is particularly notable as, according to all
previous neutron scattering studies, the Tb moments in the collinear phase are disordered.
The data in Fig.\ \ref{kscans_tb}(a) and (b) also appear to show that far from being the weaker 
peaks, at the Tb $L_3$ edge 
the $F$-type are actually considerably more intense than the $A$-type peaks. Such comparisons
should, however, be qualified by the fact that at the Tb $L_3$ edge a marked asymmetry in the intensity 
of the $\pm$q satellites was present (see Fig.\ \ref{qscans}(b)). 

\begin{figure}[t]
\centering
\includegraphics[width=0.40\textwidth,clip]{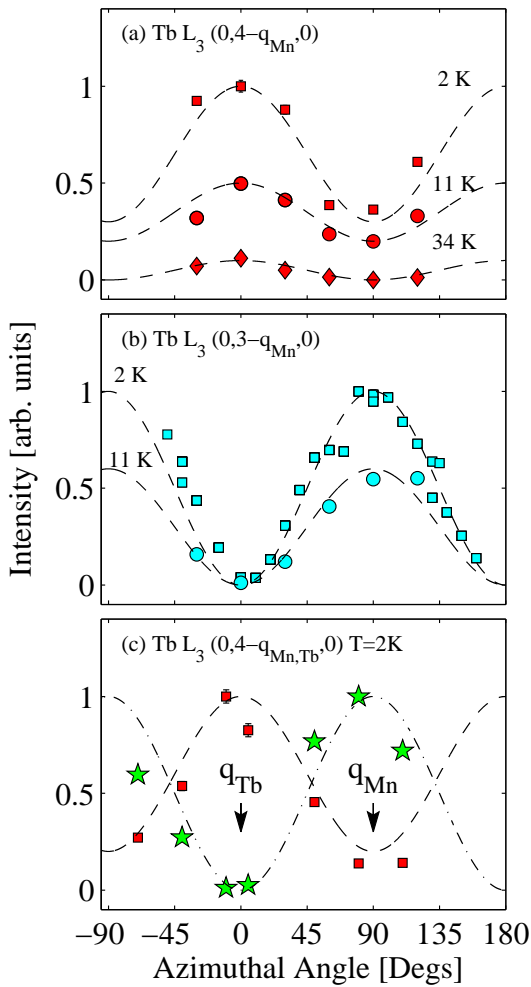}
\caption{ (Color online) 
Summary of azimuthal scans of various diffraction peaks
recorded at different temperatures and photon energies.
The data were recorded in the $\sigma-\pi^\prime$
channel and have been normalised to unity. 
The lines are guides to the eye. The azimuth $\psi=0$
corresponds to the ${\bf c}$ axis in the scattering plane.
}
\label{az_s2}
\end{figure}

\begin{figure}[t]
\centering
\includegraphics[width=0.40\textwidth,clip]{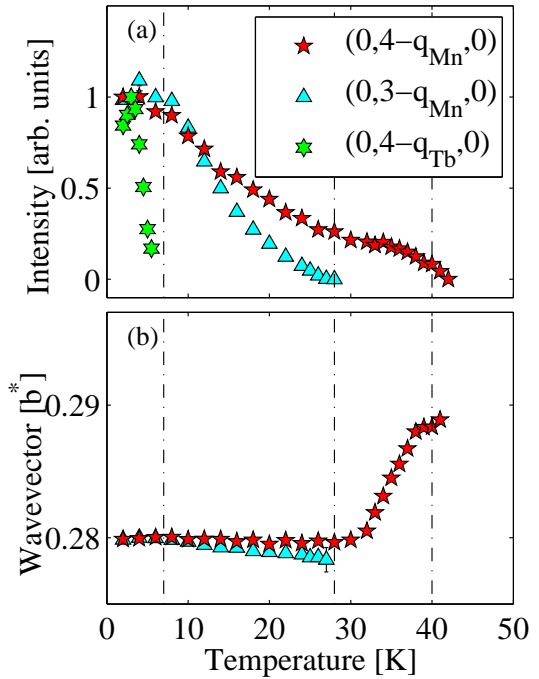}
\caption{ (Color online)
Temperature dependence of the intensity and modulation wave-vector
at the Tb $L_3$ edge recorded in 
$\sigma-\pi^\prime$. For ease of comparison, the intensities have been normalised to unity 
at low temperature.}
\label{t_2}
\end{figure}

The energy dependence of $(0,4\pm\mathrm q,0)$ is shown in Fig.\ \ref{es_tb1}(b).
Inspection of the energy scans allows us to ascribe the resonances to an
$E1-E1$ processes, which in the case of the Tb $L_3$ edge connect $2p$ to $5d$ states.
The azimuthal dependence of the $(0,4\pm\mathrm q,0)$ $\sigma$-$\pi^\prime$ intensities is shown in 
Fig.\ \ref{az_s2}(b), and can be seen to exhibit 
the same $\cos^2\psi$ dependence as found at the Mn $K$ edge (Fig.\ \ref{az_s1}(b)).
The temperature dependence of the $(0,4\pm\mathrm q,0)$ intensities and wave-vectors are shown in Fig.\ \ref{t_2} where 
we find that they follow exactly the temperature dependence of the XRS observed at the Mn $K$ edge.

From these observations, we can immediately 
deduce that in the interval $T_{N2}\leq T \leq T_{N1}$, in other words  
well above any ordering of the Tb$^{3+}$ dipole moments, the 
$5d$ states at the Tb sites are polarized by ordering of the Mn magnetic moments and with a 
modulation wave-vector identical to that of the Mn sublattice.
(In this context we note that induced polarisations
have been reported on the anion species of several actinide materials \cite{mannix2001}.)
This of course does not answer the question of the nature of the 
polarization that is created at the Tb sites, and hence the origin of the scattering process which
gives rise to the resonances. We consider the various possibilities in Sec.\ \ref{sec:ana}.

\subsubsection{Ferroelectric/cycloidal phase: Mn K edge}
\label{sec:cy_mn}

\begin{figure}[t]
\centering
\includegraphics[width=0.40\textwidth,bb=141 114 430 727,clip]{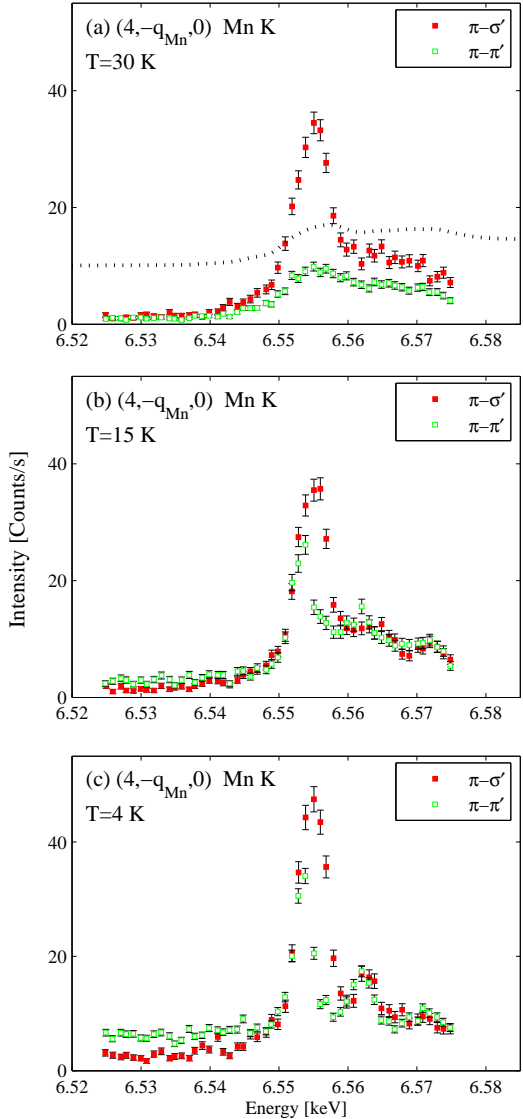}
\caption{ (Color online)
Energy dependence of XRS at the (4,-q$_{Mn}$,0) satellite in the vicinity of the Mn $K$ edge 
collected using a normal beam geometry with $\pi$ incident photons on ID20. 
}
\label{es_mn_pi}
\end{figure}

\begin{figure}
\centering
\includegraphics[width=0.40\textwidth]{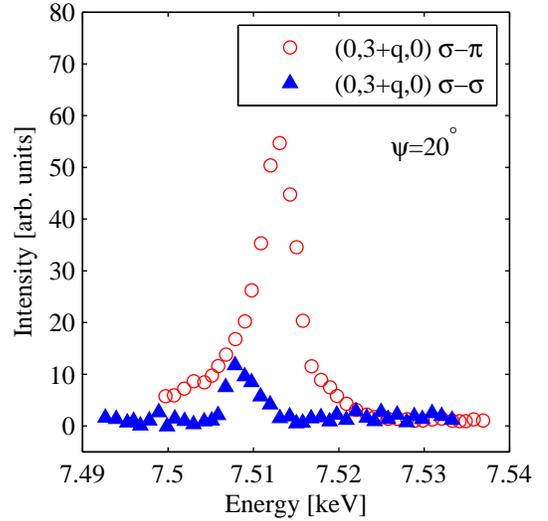}
\caption{ (Color online)
Energy dependence of XRS at the (0,3-q$_{Mn}$,0) satellite in the vicinity of the Tb $L_3$ edge 
recorded on ID20. The data were taken at a temperature of $T$=14 K, and the azimuth was set to be  
$\psi=20^\circ$, {\it i.e.} close to the minimum in the intensity of the response in the 
$\sigma-\pi^\prime$ channel (Fig.\ \protect\ref{az_s2}(b)).
}
\label{es_kodd}
\end{figure}

It is of considerable interest to follow the XRS at the Mn $K$ edge 
on entering the ferroelectric phase. In the model proposed by Kenzelmann {\it et al.},
ferroelectricity arises when the Mn sublattice makes the transition  
from a collinear to non-collinear structure, removing the inversion centre at the Mn site. 
In principle, the loss
of inversion symmetry may be detected directly in XRS experiments as the opening of new 
$E1-E2$ scattering channels in the region of the pre-edge\cite{marri2004,dimatteo2005,dimatteo2006,lovesey2005}.

The most complete data set that we have on changes to the XRS spectra at the Mn $K$ edge on entering the
ferroelectric phase 
was taken with incident $\pi$ polarized photons. The temperature dependence
of the resonant line shape of the $F$-type satellite 
$(4,-\mathrm q,0)$ is summarised in Fig.\ \ref{es_mn_pi}.
In the collinear phase at 30 K the scattering is predominantly in the rotated $\pi-\sigma^\prime$
channel, with a smooth, weaker response in the unrotated channel, most likely associated with the onset
of fluorescence at the edge. On cooling into the ferroelectric phase a dramatic change occurs in the
unrotated channel with a peak appearing displaced to lower energy by 2 eV from the peak in the response
in the rotated channel. With further cooling the intensity of the unrotated component increases until it becomes
comparable in magnitude to the rotated one. In addition a feature appears above the edge in both channels.

Inspection of Eq.\ \ref{eq:E1res} reveals that for $\pi$ polarized incident photons, $E1-E1$ XRMS scattering from
a longitudinally modulated magnetic structure polarized along ${\bf b}$ should occur in the  $\pi-\pi^\prime$
channel only. This is exactly the opposite of what is shown in Fig.\ \ref{es_mn_pi}(a), where the predominant 
scattering occurs in the $\pi-\sigma^\prime$ channel. Further consideration of Eq.\ \ref{eq:E1res} indicates
that, if the peak is to be ascribed to XRMS, then the moments in the collinear phase are instead polarized along 
${\bf a}$ or ${\bf c}$. Consistency with the data on the (0,4$\pm$q,0) satellites (Sec.\ \ref{sec:co_mn}) 
can be achieved when we recall that
azimuthal scans at this peak position suggested moments polarized along ${\bf c}$. 
The appearance of significant intensity in the unrotated channel on entering the ferroelectric phase
(Fig.\ \ref{es_mn_pi}(b)) might then further be explained, for an $E1-E1$ XRMS process, 
by the development of a component of the 
magnetization along the ${\bf b}$ axis.
However, this interpretation would seem to be untenable as the intensity in the unrotated channel 
peaks 2 eV below that in the rotated one, indicating that the former is not pure $E1-E1$.
Here the caveats raised in Sec.\ \ref{sec:co_mn} and \ref{sec:co_tb}  
about interpreting the $F$-type peaks as XRMS  should also be recalled. We do not observe
any $C$-type XRS peaks at positions such as (0,3$\pm$q,0) at the Mn $K$ edge,
which indicates that these Mn magnetic domains are absent in our sample, in agreement with
our NRXMS results.   

\subsubsection{Ferroelectric/cycloidal phase: Tb L$_{3}$ edge}
\label{sec:cy_tb}

Upon cooling into the cycloidal phase  $T_{N2}\leq T \leq T_{N1}$,  
a new set of $C$-type resonant satellites at $(0,k\pm\mathrm q,0)$ $k$ odd were revealed at the Tb $L_3$ edge.
These were found to be absent at the same temperature at the Mn $K$ edge and unlike the $A$-type and $F$-type 
reflections,
we do not find any charge order satellites at $2q_{Mn}$ associated with these $C$-type peaks.  
Key data on their energy dependence measured at $(0,3+\mathrm q,0)$ are shown in Fig.\ \ref{es_kodd} where it is apparent
that the intensity appears  mostly in the
rotated $\sigma$-$\pi^\prime$ channel.
The azimuthal dependence of the $(0,3+\mathrm q,0)$ is shown in Fig.\ \ref{az_s2}(b) and Fig.\ \ref{az_s3}, where it
is seen to have a $\sin^2\psi$ dependence, in anti-phase to the azimuthal dependence found at
$(0,4+\mathrm q,0)$ shown in Fig.\ \ref{az_s1}(b) and Fig.\ \ref{az_s2}(a). At these $C$-type positions, energy
scans also reveal a weaker pre-edge feature as seen in Fig.\ \ref{es_kodd}, 
which is scattered with $\sigma$-$\sigma^\prime$ polarisation. The azimuthal dependence of this 
peak has a
$\cos^2\psi$ azimuthal dependence (Fig.\ \ref{az_s3}).

The temperature dependence of the $\sigma$-$\pi^\prime$ polarised intensities are shown in Fig.\ \ref{t_2}(a)
where they are seen to disappear above $T_{N2}$=28~K. Magnetic 
satellites were also identified at these wave-vectors in the neutron
scattering investigation by Kenzelmann {\it et al.}\cite{kenzelmann2005} 
and were attributed to an induced moment on the Tb ions, oriented
along the ${\bf a}$ axis, from the onset of the Mn magnetic structure 
in the cycloid phase. From Eq.\ \ref{eq:E1res} it can be seen
that the observed azimuthal dependence is consistent with a magnetic moment directed along the ${\bf a}$ axis.
Detailed analysis described in Sec.\ \ref{sec:ana} substantiates this further, but also indicates 
that the weak pre-edge feature in the unrotated channel shown in Fig.\ \ref{es_kodd} may reflect the 
contribution from an order parameter with both odd time and inversion parity. 

In summary, in the cycloidal phase  additional XRS satellites
appear at $(0,k\pm\mathrm q,0)$, $k$ odd, for photon energies around the Tb $L_3$ edge, but not 
the Mn $K$ edge. They are found predominantly in the rotated polarization channel, and 
are most probably $E1-E1$ XRMS from a polarisation
of the Tb $5d$ states  along the {\bf a} axis induced by the onset of the Mn cycloidal order.
These findings are in agreement with the model of Kenzelmann {\it et al.} for this phase.

It is of course important to reconcile these XRS observations at the Tb edge with
our failure to observe NRXMS at $(0,k\pm\mathrm q,0)$ $k$ odd, discussed 
in Sec.\ \ref{sec:nrxms}. This may be achieved if we suppose that the Tb 4f moments 
are only weakly polarised in the cycloidal phase. This combined with the highger background (Fig.\ \ref{nrss}(d))
observed in the $\sigma$-$\sigma^\prime$ channel where the peak would be expected to occur (Eq.\ \ref{eq:nrxms}) 
would then
act to render the peak unobservable.

\subsubsection{Low temperature phase: Tb L edge}
\label{sec:lt_tb}

\begin{figure}[t]
\centering
\includegraphics[width=0.35\textwidth,bb=106 145 491 750,clip]{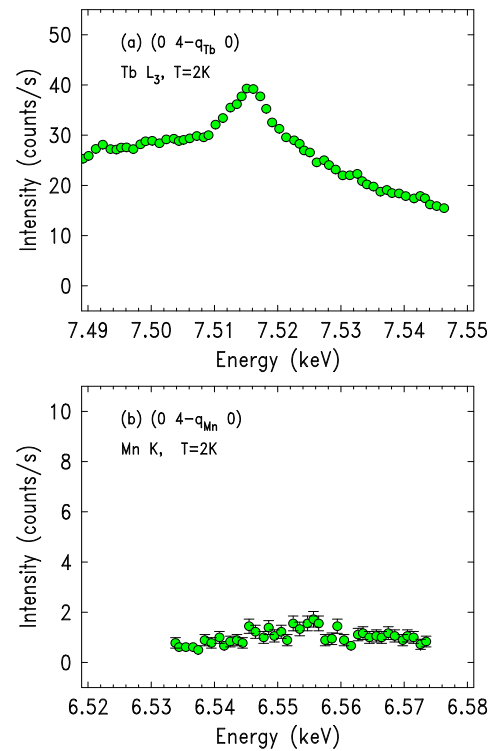}
\caption{(Color online)
(a) Energy scan taken at the Tb ordering wave-vector q$_{Tb}$=0.42 at T=2K in the 
$\sigma-\pi^\prime$ channel, where only weak XRS is observed. (b) Energy scan
taken at the Mn ordering wave-vector q$_{Mn}$=0.28 at T=2K, where no significant XRS is observed.
}
\label{qtb_2}
\end{figure}

We have additionally investigated the XRS below $T_{N3}=7$~K, where the Tb 
moments order, with the distinct wave-vector of $(0,\mathrm q_{Tb},0)$ 
with q$_{Tb}$=0.42 b$^*$. However, in this phase, energy scans taken at q$_{Tb}$ find only a
weak XRS response, as shown in Fig.\ \ref{qtb_2} (a). The peak in the energy scan at
the Tb L$_{3}$ edge is close to the maximum of the fluorescence spectrum and allows us to
associate the XRS as arising from $E1-E1$ transitions, probing the Tb $5d$ states. For completeness, 
we also show in Fig.\ \ref{qtb_2} (b) an energy scan of q$_{Tb}$ taken at the Mn K-edge, where no significant
XRS can be identified. This implies that, within the accuracy of our measurements, we find no evidence for an
induced polarisation of the Mn ions from the onset of the Tb order.  

\begin{figure}[t]
\centering
\includegraphics[width=0.35\textwidth,bb=87 145 481 688,clip]{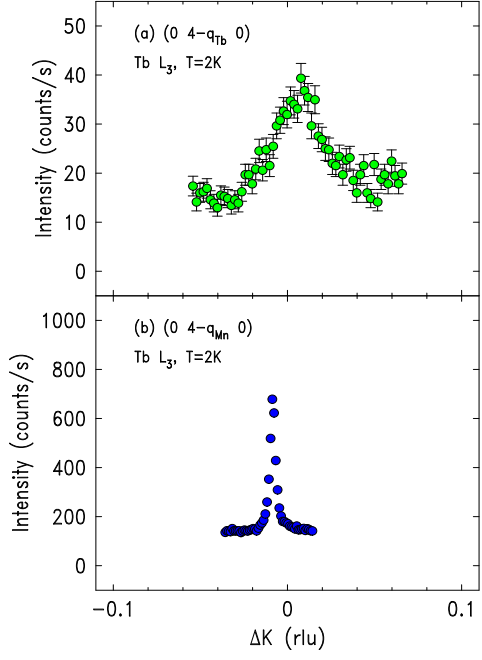}
\caption{ (Color online) 
(a) k-scan taken at the Tb ordering wave-vector q$_{Tb}$=0.42 at T=2K in the 
$\sigma-\pi^\prime$ channel. The broad width of the scattering indicates that
underlying magnetic order is only short ranged ordered at this temperature. (b) k-scan
taken at the Mn ordering wave-vector q$_{Mn}$=0.28 at T=2K indicating that the induced polarisation
on the Tb ions are longed ranged order at this position.
}
\label{qtb_1}
\end{figure}

A scan of the wavevector transfer parallel to [0$k$0] through  
q$_{Tb}$ provide us with more information about the weakness of the XRS, Fig.\ \ref{qtb_1} (a).
This scan reveals that the peaks associated with q$_{Tb}$  are very broad in reciprocal space, even 
at the lowest temperature measured of 2 K and are therefore consistent with only short range order. 
Presumably, this effect arises from the competition of magnetic order
at this position with the induced Tb ordering we identify 
at $(0,k\pm\mathrm q,0)$ $k$ odd. Indeed, it is intriguing that these XRS investigations
appear to find that the major polarisation of the Tb valence electrons arises from the Mn $3d$ order
and not from the ordering of the Tb $4f$ ions. The short ranged order at q$_{Tb}$ has also been reported
in forgoing neutron investigations \cite{blasco2000, kenzelmann2005}. It may be of significance that
the coupling of Mn and Tb moments have been invoked as higher order terms in the nearest-neighbour and 
next-nearest-neighbor(NN-NNN) model to explain the 
emergence of the incommensurate spin structure in the heavy rare-earth ReMnO$_3$ systems
\cite{kajimoto2004, kimura2003b}.

The azimuthal dependence of $(0,4-\mathrm q_{Tb},0)$ 
is shown in Fig.\ \ref{az_s2}(c), and has 
identical azimuthal symmetry as those recorded at $(0,k\pm\mathrm q,0)$ $k$ odd
as shown in Fig.\ \ref{az_s1}(b). This comparison 
indicates that the low temperature ordered Tb magnetic moments are orientated along the ${\bf a}$ axis.
The temperature dependence of the intensity of  $(0,4-\mathrm q_{Tb},0)$ 
is shown in Fig.\ \ref{t_2}(b), where
it disappears above $T_{N3}$ in agreement with neutron 
scattering results. 

\begin{figure}[t]
\centering
\includegraphics[width=0.45\textwidth,clip]{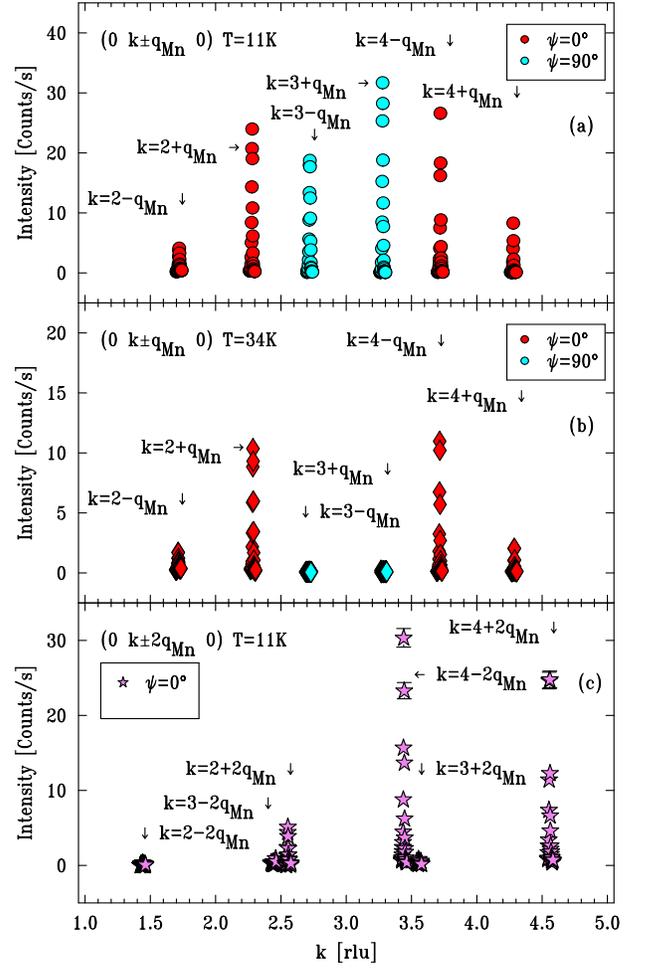}
\caption{ (Color online) 
Summary of scans of the wavevector transfer along [0,$k$,0].
(a) and (b) XRS data around the primary satellite positions 
$(0,k\pm \mathrm q_{Mn},0)$ at the Tb $L_3$ edge for 
azimuthal angles of $\psi$= 0 and 90$^\circ$ in the $\sigma-\pi^\prime$ channel. 
(c) Non-resonant, charge satellites at $(0,k\pm 2\mathrm q_{Mn},0)$
in the $\sigma-\sigma^\prime$ channel.}
\label{qscans}
\end{figure}

\subsubsection{Charge satellites}
\label{sec:2q} 

We have also investigated the satellites which occur 
at 2q$_{Mn}$ in TbMnO$_3$. We find that these satellites around a given
A-type magnetic wave-vector such as $(0,4-\mathrm q,1)$ 
do not resonate at either the Tb or Mn absorptions edges. Of significance
is the fact 
that we also find 2q$_{Mn}$ satellites associated with $(0,4-\mathrm q,0)$.
This result suggest that the charge degrees of freedom also 
couple to the lattice causing the charge order satellites that we observe
and indicating an intricate interplay of magnetic and charge 
lattice coupling in this material. We also found that these 2q$_{Mn}$ peaks do
not resonate at either absorption edge. Key data on the 
wave-vector scans are shown in Fig.\ \ref{qscans} for the resonant q$_{Mn}$  
and non-resonant 2q$_{Mn}$ satellites. It is also noteworthy that we do not find
any evidence for charge ordered satellites associated with the $C$-type satellites
such as $(0,3-\mathrm q,0)$.  

\subsection{Analysis and Discussion}
\label{sec:ana}
 
To obtain a deeper understanding of our resonant scattering investigations,
we have performed {\it ab initio} electronic structure calculations using the 
$FDMNES$ code \cite{joly2001} which allowed us to evaluate 
the XRS spectra, including possible contributions from $E1-E1$, $E1-E2$ and 
$E2-E2$ processes. 

It is important to point out that $E1-E1$ XRS can, in most cases, be easily identified
because the energy maximum occurs very close to that in the fluorescence spectrum. However,
XRS features which occur at energies corresponding to the pre-edge threshold may arise from
$E1-E1$,$E2-E2$ and $E1-E2$ events and cannot be simply interpreted from their position in energy.



\subsubsection{$FDMNES$ code}

\begin{figure}[t]
\centering
\includegraphics[width=0.40\textwidth,clip]{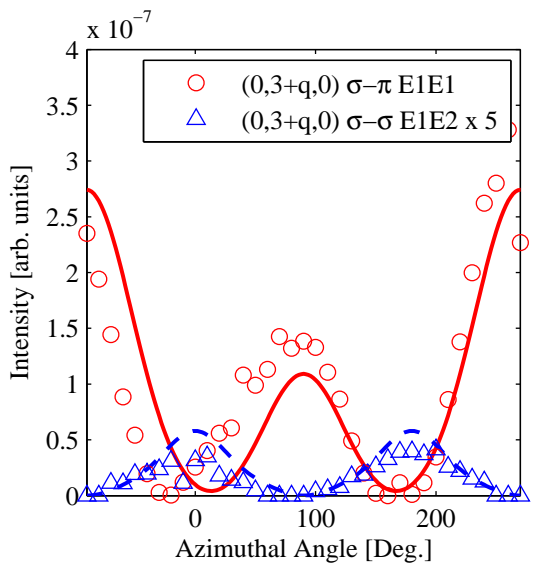}
\caption{ (Color online) 
Azimuthal dependence of XRS at the (0,3-q$_{Mn}$,0) satellite in the vicinity of the Tb $L_3$ edge 
recorded on ID20. The $\sigma-\pi$ data were taken at an energy corresponding to an $E1-E1$ event 
around 7.520 keV, while the $\sigma-\sigma$ data were taken 7 eV below this energy.
The lines have been calculated using the $FDMNES$ package.
In the unrotated channel, the main contribution is calculated to be from $E1-E1$ XRMS associated with 
a splitting of the Tb $5d$ bands induced by the cycloidal order on the Mn sublattice. For the unrotated 
channel the weak pre-edge peak is calculated to be $E1-E2$, arising from an anapole, {\it i.e.} a 
multipole which is odd with respect to both time and parity. The data in the unrotated channel have been
multiplied by a factor of five.
}
\label{az_s3}
\end{figure}

$FDMNES$ is a software package that allows the calculation of the intensity of the diffracted reflections 
in the vicinity of absorption edges of the elements present in the material of interest.
The code may use various methods to calculate the excited states probed during 
the resonant process and in our calculations, we employed the multiple scattering theory. The 
final states are calculated in a fully relativistic way, including the spin-orbit contribution. Then the 
matrix elements governing the transition process between the initial and final states are calculated considering 
the polarization conditions of the incoming and outgoing photon. This photon electric field is expanded in order 
to calculate both the  electric dipole ($E1$) and electric quadrupole ($E2$) contributions and consequently 
for the resonant process the $E1-E1$,
$E1-E2$ and $E2-E2$ components. At the end the structure factors are calculated and an expansion in a spherical basis 
is performed, so at the various diffraction satellites the different contributions coming from all the order of 
scattering can be separately provided. In this way one obtains the monopole ($F{^0}$), dipole ($F^1$) 
and quadrupole ($F{^2}$) $E1-E1$ 
contributions, the time even ($+$) 
and time odd ($-$) dipole ($F{^1\pm}$), quadrupole ($F{^2\pm}$) and octupole ($F{^3\pm}$) $E1-E2$ contributions 
and the monopole,
dipole, quadrupole, octupole and hexadecapole $E2-E2$ contributions. For the $E1-E2$ component the non magnetic 
and magnetic parts can be given separately, so it is 
possible to precisely deduce which terms in particular contribute to the XRS intensity. 


In practice one has to provide to the code the crystallographic structure and a guess of the electronic structure, 
including the spin polarization. From this the Dirac equation is solved. At low temperature neutron diffraction
experiments have shown that the crystallographic structure does not change by very much \cite{blasco2000} from the
high temperature \emph{Pbnm} structure. The main problem in the context of the present study is that even minute changes in
crystal structure greatly effect the XRS. We chose to keep the $Pbnm$ 
structure we built a supercell, four times bigger along the ${\bf b}$ direction to mimic 
the magnetic modulated structure of Kenzelmann et al. In this way the corresponding new diffraction 
vectors $(h,k\pm q,l)$ with q = 0.25 b$^*$ is not far from the incommensurate one measured. In order to model
the data in the cycloidal phase below $T_{N2}$, we kept the same diffraction vector with q = 0.25 b$^*$ and
allowed the magnetic moment to modulate in the ${\bf b-c}$ plane 
as described by Kenzelmann {\it et al}.\cite{kenzelmann2005}. 
Making these approximations we 
used $FDMNES$  to calculate the XRS intensities at the wave-vectors corresponding to $A$-type $(0,4\pm\mathrm q,1)$,
$F$-type $(0,4\pm\mathrm q,0)$ and C-type $(0,3\pm\mathrm q,0)$ order, 
which were extensively investigated in our experiments.
In all of the following discussion, intensities
below a threshold of $1 \times 10^{-6} r_0^2$ were taken to be zero. 
This is a reasonable approximation as from the $FDMNES$ calculations, the intensity
at $(0,4\pm\mathrm q,1)$ at the Mn $K$-edge is $5 \times10^{-4} r_0^2$. 
The corresponding intensity measured in our experiments at ID20
at this wave-vector was about 100 cts/second, so that intensities of the order 500 times below 
this are beyond the sensitivity of our measurements.
For the collinear phase, the calculated intensities of all possible multipolar contributions were 
found to be negligibly small at the Tb $L_3$ edge, while 
the results for the Mn $K$ edge in this phase are given in Tab. \ref{tab:mn_col}. 
Corresponding calculated intensities
for the cycloidal phase are given in  Tab. \ref{tab:mn_cyc}  and \ref{tab:tb_cyc}. 
The $E1-E1$ intensities are given 10 eV above the Fermi level (0 eV), while the $E1-E2$ and $E2-E2$ 
intensities are given at a pre-edge value taken
5 eV below the Fermi level.

\begin{figure*}[t]
\centering
\includegraphics[width=0.80\textwidth,clip]{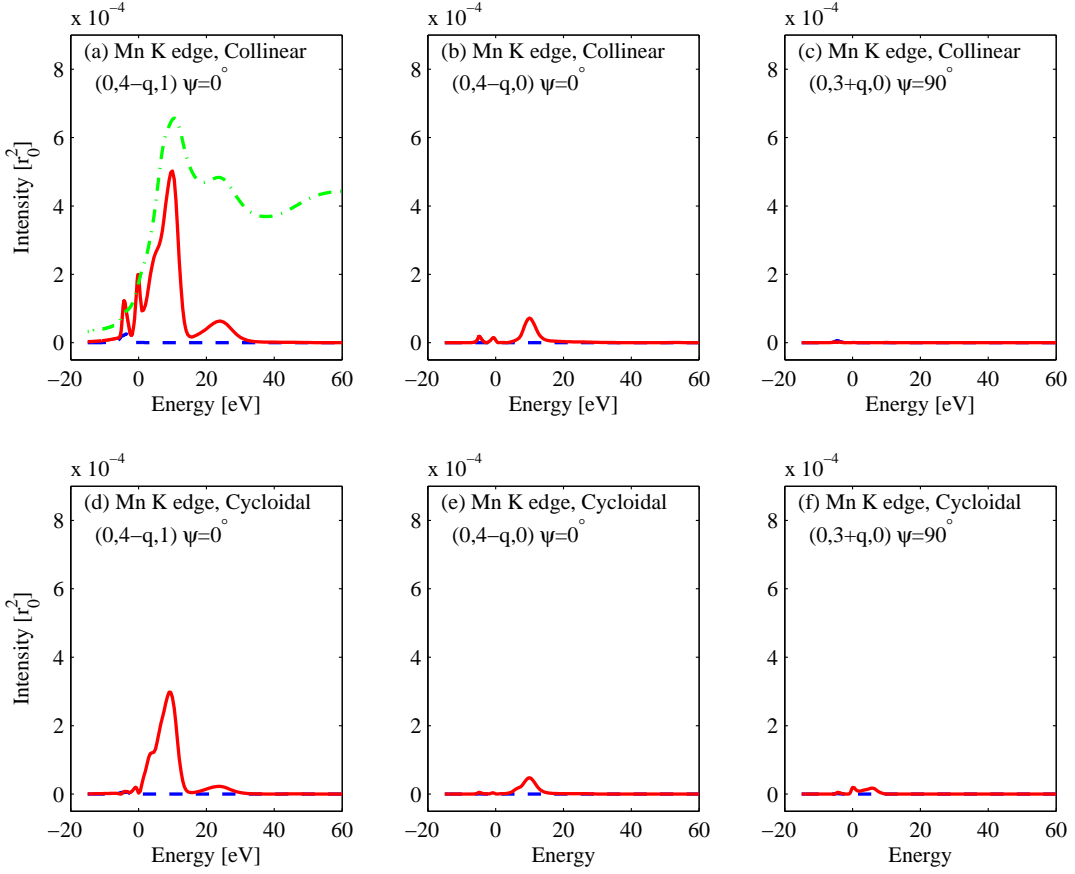}
\caption{ (Color online) 
Representative results of $FDMNES$ calculations for TbMnO$_3$ 
in the vicinity of the Mn $K$ edge for the collinear ((a)-(c)) and 
cycloidal ((d)-(f)) phases. The calculated x-ray absorption near edge structure (XANES) is given by the dot-dashed line
in (a). The figure allows for a comparison of the calculated intensities
of the $A$-type ((a) and (d)), $F$-type ((b) and (e)) and $C$-type ((c) and (f)) studied in our experiments. 
The $\sigma-\pi^\prime$ and $\sigma-\sigma^\prime$ channels are represented by the solid red and dashed blue lines, 
respectively.
}
\label{mn_yves}
\end{figure*}

\subsubsection{Collinear phase}

We first consider our results for the collinear phase, $T_{N2}<T<T_{N1}$. 
Here our XRS data reveals two sets of satellites: $A$-type, such as $(0,k\pm\mathrm q,1)$; 
and $F$-type $(0,k\pm\mathrm q,0)$, both with $k$ even. 
These two sets of satellites showed resonant behaviour 
at both the Mn $K$ and the Tb $L_{3}$ edges. 
The results of the $FDMNES$ calculation at the Mn $K$-edge, using the $\bf b$-axis 
modulated magnetic structure of Kenzelmann {\it et al.}\cite{kenzelmann2005} and the $Pbnm$ structure, 
are shown in Fig.\ 
\ref{mn_yves}(a)-(c) for the $A$-type, $F$-type and $C$-type structures, respectively, 
while the decomposition of the intensity into 
its multipolar components is given in Tab. \ref{tab:mn_col}.
The calculations predict finite $E1-E1$ $F^1$ at the $A$-type wavevectors, of the order $5 \times 10^{-4} r_0^2$, and
that the $E2-E2$ intensities are negligibly small. The pre-edge peaks shown in Fig.\  \ref{mn_yves} (a), are all
$E1-E1$ in origin. The $E1-E2$ multipoles are negligible in these calculations, since the Mn ions occupy sites with
inversion symmetry in the $Pbnm$ structure. Interestingly, the code also predicts some finite $E1-E1$
XRMS arising from the $F^1$ multipole at the $F$-type $(0,4\pm\mathrm q,0)$ wave-vector, 
which is an order of magnitude weaker than
that calculated at $(0,4\pm\mathrm q,1)$. This result is in contradiction with our x-ray experiments, 
where we find similar XRS intensities
for the $A$- and $F$-type peaks at the Mn $K$-edge as shown in Fig.\  \ref{kscans_mn} (b) and (d), 
but adds weight to the idea
idea that the XRS observed at the $F$-type peaks is not predominantly magnetic in origin. 
Shifting the Mn ions 
from their nominal positions in the calculations 
had the effect of causing larger scattering at the $F$-type wave-vector, but yielded 
azimuthal dependences inconsistent with our experimental data.
The $FDMNES$ calculations predict no significant XRS at the $C$-type 
peaks as shown in Fig.\ \ref{mn_yves} (c), in agreement with our experimental observations in the collinear phase.

Calculations using $FDMNES$ of the XRS in the collinear phase at Tb $L_3$ edge were also performed. 
The calculated XRS spectra are shown in Fig.\ \ref{tb_yves} (a)-(c) where they are seen to be 
negligibly small in the sense described above. Thus the
calculations do not reproduce our experimental results, where we observed significant XRS intensities 
at both the $A$- and
$F$-type peaks in the collinear phase at the Tb $L_3$ edge. A possible explanation for this discrepancy 
is that the $FDMNES$ calculations do not
predict a magnetic polarisation on the Tb $5d$ states induced from the Mn collinear magnetic structure. 
However, it is noteworthy that the calculations
do identify a polarisation on the Tb ions in the cycloid phase (discussed below) for the $C$-type reflections, 
induced by the Mn cycloid
magnetic structure. This induced polarisation is in agreement with our experimental results and 
the neutron scattering results of Kenzelmann {\it et al}. 
In the light of this fact, another plausible scenario is that the XRS observed at the Tb edge for 
both the $A$- and $F$-type peaks are not 
predominantly
magnetic in origin. Instead they might arise from a charge multipole order parameter, which do not 
appear with measurable intensity in the 
calculations because of assumptions made in the model, such as the restriction to
the $Pbnm$ space group. Neither experiment nor calculations yet provide a conclusive answer to this issue. 

\subsubsection{Cycloidal phase}

For the cycloidal phase at the Mn $K$ edge, the calculated XRS spectra are shown in Fig.\ \ref{mn_yves}(d)-(f)
and the corresponding multipole intensities
are summarised in Tab.\ \ref{tab:mn_cyc}. These calculation predict significant XRMS 
intensity ($1 \times 10^{-3}$ $r_{0}^{2}$)
for the $A$-type and weaker intensities for the $F$-type ($5 \times 10^{-5}$ $r_{0}^{2}$) 
and $C$-type ($1 \times10^{-6}$ $r_{0}^{2}$) 
peaks, all of which are magnetic dipole in origin, $E1-E1$ $F^1$. As with the comparison in the collinear phase, 
there is considerable disagreement between experiment and calculation in the ratio of intensities at the 
$A$- and $F$-type peaks. Although the calculations indicate
that the $F$-type intensities are around two orders of magnitude smaller, our experiments find similar XRS 
intensities for the two peaks. Again, this may be
an indication that the XRS we observe at the $F$-type peaks is not magnetic in origin. The calculations 
predict a very weak XRS at the $C$-type wave-vectors, in disagreement with our experimental results where these
peaks were found to be absent. 
However, this weak intensity may be at the limit of our 
experimental sensitivity. Again, as in the collinear phase at the Mn $K$ edge, all $E1-E2$ XRS is calculated to
be negligibly small
at this absorption edge because the Mn ions are assumed to occupy positions of centre inversion symmetry in the $Pbmn$ 
space group.           
   
For the ferroelectric phase, $T_{N3}<T<T_{N2}$, the most striking feature of our XRS data 
is the appearance of $C$-type, $(0,k\pm\mathrm q,0)$, $k$ odd, satellites for photon energies around the Tb $L_3$ edge.
In this phase, the XRS cross-section at the Tb $L_{3}$ 
was calculated by inputing the cycloidal magnetic structure. 
Here the $FDMNES$ code predicts a measurable scattering intensity due an 
$E1-E1$ XRMS process in the $\sigma-\pi^\prime$ channel of $4 \times 10^{-4}$. Remarkably the approximately $\sin^2\psi$ 
azimuthal dependence predicted in this channel appears to account for the data shown in 
Fig.\ \ref{az_s2}(b). The $FDMNES$ code actually reveals that  
scattering at this wavevector in the rotated channel arises from a small magnetic net polarization
of the Tb $5d$ bands along the ${\bf a}$-axis induced by the onset of cycloidal magnetic order on the Mn sublattice. 
In contrast, only weak XRS intensities were calculated at corresponding wave-vectors at the Mn $K$
edge.  

\begin{figure*}[t]
\centering
\includegraphics[width=0.80\textwidth,clip]{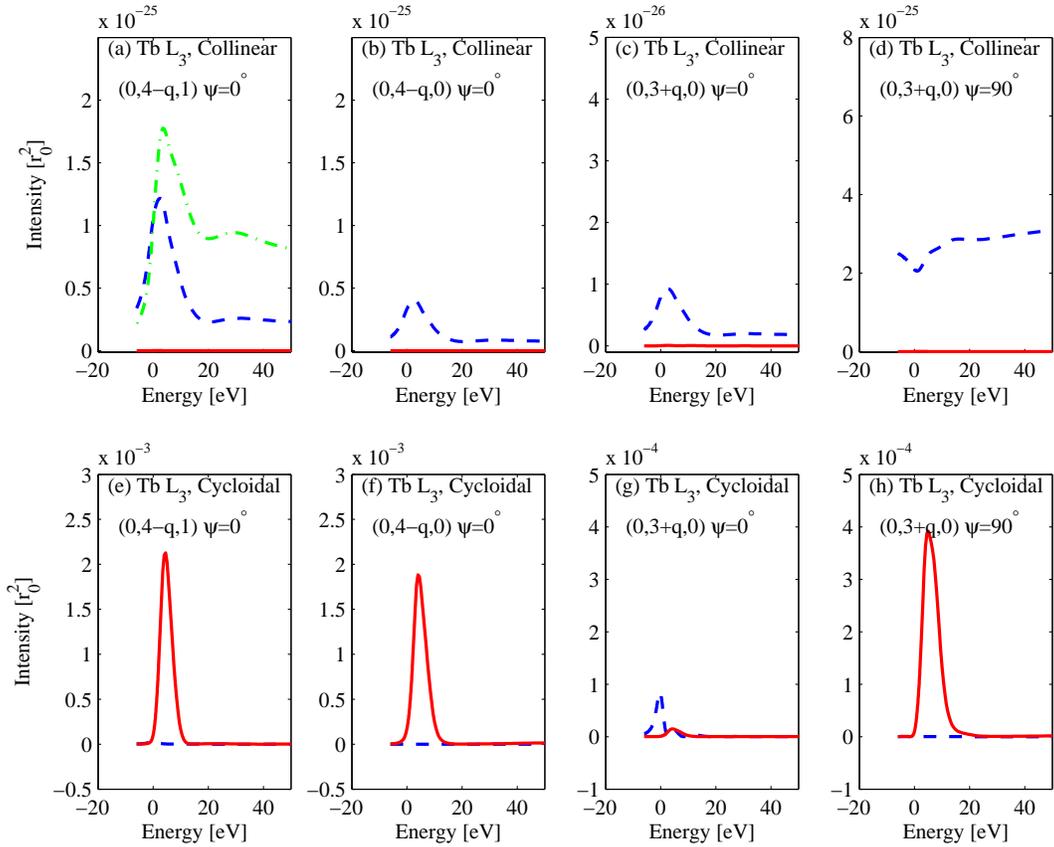}
\caption{ (Color online) 
Representative results of $FDMNES$ calculations for TbMnO$_3$ 
in the vicinity of the Tb $L_3$ edge for the collinear ((a)-(d)) and 
cycloidal ((e)-(h)) phases. The calculated x-ray absorption near edge structure (XANES) is given by the dot-dashed line
in (a). The figure allows for a comparison of the calculated intensities
of the $A$-type ((a) and (e)), $F$-type ((b) and (f)) and $C$-type ((c), (d), (g) and (h)) studied in our experiments. 
The $\sigma-\pi^\prime$ and $\sigma-\sigma^\prime$ channels are represented by the solid red and dashed blue lines, 
respectively.
}
\label{tb_yves}
\end{figure*}

The azimuthal symmetry at $(0,k\pm\mathrm q,0)$, $k$ odd, is actually more complex 
than the simple $\sin^2\psi$ form suggested by the data in Fig.\ \ref{az_s2}(b) for the $\sigma-\pi^\prime$ channel. 
A much more extensive set of measurements were performed with the 
results shown in Fig.\ \ref{az_s3}. Not only is the azimuthal dependence clearly more complex, with 
one-fold rather than two-fold symmetry, but in addition these
experiments revealed the existence of a weak component to the scattering 
in the unrotated ($\sigma-\sigma^\prime$) channel. Most
importantly, this component has an azimuthal dependence that peaks in anti-phase with
the rotated ($\sigma-\pi^\prime$) component (Fig.\ \ref{az_s3}), and resonates 
7 eV below the position of the $E1-E1$ dipole resonance (Fig.\ \ref{es_kodd}). The latter
fact indicates that it arises either from an $E2-E2$ or from an $E1-E2$ process. 
Calculations using the $FDMNES$ code indicate that, although
the $E2-E2$ and $E1-E2$ spectra have similar azimuthal 
dependencies, the $\sigma-\sigma^\prime$ signal actually
arises mostly from an $E1-E2$ event at this position (Tab. \ref{tab:mn_cyc}). 
Specifically, this component is consistent with the existence of a finite $F^{1-}$ term,
a time-odd, parity-odd multipole, which is also known as the anapole. 
The calculations indicate that the $E2-E2$ contributions
are very small at this position because the $FDMNES$ band structure calculations
predict that the induced splitting is largly in the $5\emph{d}$ and not the
$4\emph{f}$ band.

At this stage it is not possible to make a definitive conclusion regarding the observation of an anapolar 
contribution to the XRS in multiferroic TbMnO$_3$. For one thing, the agreement between the data and 
the calculations based on the $FDMNES$ code are qualitative only. While the code does a good job of accounting for the 
azimuthal dependences (Fig.\ \ref{az_s3}) and spectra (Fig.\ \ref{es_kodd} and Fig.\ \ref{tb_yves}(g)-(h)), 
it overestimates the relative intensity of the ($\sigma-\sigma^\prime$) by a factor of roughly $5$. 
However, it is probably reasonable to assert that, given the complex nature of the calculations, that the agreement is 
about as good as might be achievable. 

We reiterate here that the FDMNES calculations predicting an anapolar contribution
at the C-type reflections assume a $Pbnm$ space group. For this structure the Tb ions lack a centre of inversion and even
in the absence of any ferroelectric order an anapole moment on the Tb sites is allowed by symmetry. In the FDMNES 
calculations the anapole contribution presumably is induced by the non-collinearity of the Mn moments in the cycloidal
phase. Whether additional mechanisims contribute to the possible anapole moment, such as small shifts of the Tb
positions, remains an open question.

Another explanation for the discrepancy between the calculated and experimental
polarisation dependence at the $C$-type reflections 
could be that, as we have already evoked for the $A$ and $F$ reflections
observed at the Tb edge in the collinear phase: the $C$-type peaks might also have a charge multipole contribution in
the cycloid phase. This could have the effect of increasing our measured $E1-E1$ XRS and thus making the experimental
polarisation ratio larger than the calculated one. In this respect we are unable to draw a definitive conclusion.
We also note a further weakness with the analysis. At present the 
$FDMNES$ code can only deal with one magnetic ion at a time.
In addition to the cycloidal magnetic order on the Mn sublattice, the model of 
Kenzelmann {\it et al.}\cite{kenzelmann2005} has 
1.0 $\mu_B$ on the Tb sublattice polarized along the ${\bf a}$-axis. Including just this latter component in the 
$FDMNES$ code revealed unsurprisingly that at $(0,k\pm\mathrm q,0)$, $k$ odd, the $E2-E2$ contribution is larger 
than the one from $E1-E2$. However, this also 
introduced significant intensity at $(0,4\pm\mathrm q,0)$ with an incorrect
azimuthal dependence and we therefore consider it to be less consistent with our experimental data. 

\subsection{Summary}
\label{sec:sum}
We have carried out a comprehensive series of x-ray scattering experiments on multiferroic TbMnO$_{3}$
using both non-resonant and resonant techniques.
In conceiving of the experiments our hope was to be 
able to exploit the exquisite (and unique) 
ability of XRS to reveal and differentiate between possible multipolar
order parameters. In particular, the transition from the collinear to the cyloidal/ferroelectric state in 
TbMnO$_{3}$, with  the concomitant loss of inversion symmetry, 
should, at some level of sensitivity, be reflected by new features 
appearing below the transition arising from mixed $E1-E2$ processes. Evidence for such changes have indeed been 
found at both the Mn $K$ (Fig.\ \ref{es_mn_pi}) and Tb $L$ (Fig.\ \ref{es_kodd}) edges. However, 
the job of relating these observed changes to a microscopic description of the active multipoles is a 
considerable challenge, and one that we have only gone part way to completing here. 
Our hope is that our data will serve to stimulate further theory and experiments in this class of materials. 

More specifically, aided by NRXMS, which allowed us to characterise
the magnetic domain structure within the sample volume probed by the x-ray beam,  
our experiments appear to have uncovered two previously unknown order
parameters in this material. The NRXMS investigation has determined unambigously,
that our sample has a single $A$-type magnetic domain only. Nethertheless, XRS scattering
is identified at $F$-type and $C$-type wavevectors, neither of which can be reconciled
with XRMS from the nominal Mn magnetic order. Since the NRXMS results appear to 
exclude the $E1-E1$ $F$-type
XRS observed at the Mn $K$-edge from being magnetic in origin, we propose that it most likely
arises from an induced charge multipole order parameter. If this is correct, then of crucial
importance are the charge order satellites observed at $2q_{Mn}$ about these $F$-type satellites, which suggest an
intricate magnetic and charge order coupling in the material. We have attempted to further analyse
the XRS results by implementing the $FDMNES$ code, with varying degrees of success. Perhaps the most
noteworthy outcome of these calculations is the identification of significant $E1-E2$ events
in the Tb $L_{3}$ pre-edge of the XRS at $C$-type wavevectors in the magnetic cycloidal phase.
Although our results at this stage should be considered as indicative only, our experiments 
may provide evidence for the existence of a novel type of anapolar oder parameter 
in the rare-earth manganite class of multiferroic compounds. 

\section*{Acknowledgments}
\noindent
We would like to thank S. Lovesey, K. Knight, C. Detlefs, M. Kenzelmann, P. Hatton and A. Wills for helpful discussions.
Work in Oxford was supported by  the EPSRC and in London by a
Wolfson Royal Society Research Merit Award and the EPSRC.

\appendix

\section{$FDMNES$ Multipole Calculations}


\begin{table*}
\begin{center}
\begin{ruledtabular}
\begin{tabular}{|c|c|c|c|c|c|c|c|}
\multicolumn{8}{|c|}{Collinear Phase Mn $K$ edge}\\
\hline
Transition &  Multipole &  $(0,4\pm\mathrm q,1)$ &  $(0,4\pm\mathrm q,1)$ &  $(0,4\pm\mathrm q,0)$  & $(0,4\pm\mathrm q,0)$ &  $(0,3\pm\mathrm q,0)$ & $(0,3\pm\mathrm q,0)$ \\
& &  $\sigma\sigma^\prime$ & $\sigma\pi^\prime$  &  $\sigma\sigma^\prime$ & $\sigma\pi^\prime$  & $\sigma\sigma^\prime$ & $\sigma\pi^\prime$ \\ 
\hline
\multirow{1}{*}{E1-E1} 
 & $F^{1}$ & 0 & $6\times 10^{-4}$ & 0 & $6 \times 10^{-5}$ & 0 & 0 \\
\hline
\multirow{2}{*}{E2-E2} & $F^{0}$ & 0 & 0 & 0 & 0 & 0 & 0 \\
 & $F^{1}$ & 0 & 0 & 0 & 0 & 0 & 0 \\
 & $F^{3}$ & 0 & 0 & 0 & 0 & 0 & 0 \\
\end{tabular}
\end{ruledtabular}
\caption{The results of $FDMNES$ calculations of XRS multipole intensities at 
the Mn $K$ edge, in units of electrons$^{2}$, in the collinear phase. All 
calculations were undertaken for azimuth $\psi=0$. The $E1-E2$ multipoles are 
negligible in the calculations because the Mn ions are located at cites 
with centre inversion symmetry in the $Pbnm$ space group. The even order $E1-E1$ and $E2-E2$ and the time-even $E1-E2$ terms are explicitly zero
because they cannot arise from our magnetic structure. In the table zero means the 
calculated intensities were less than $1 \times 10^{-6}$ i.e. very small, not zero}
\label{tab:mn_col}
\end{center}
\end{table*}

\begin{table*}
\begin{center}
\begin{ruledtabular}
\begin{tabular}{|c|c|c|c|c|c|c|c|}
\multicolumn{8}{|c|}{Cycloidal Phase Mn $K$ edge}\\
\hline
Transition &  Multipole &  $(0,4\pm\mathrm q,1)$ &  $(0,4\pm\mathrm q,1)$ &  $(0,4\pm\mathrm q,0)$  & $(0,4\pm\mathrm q,0)$ &  $(0,3\pm\mathrm q,0)$ & $(0,3\pm\mathrm q,0)$ \\
& &  $\sigma\sigma^\prime$ & $\sigma\pi^\prime$  &  $\sigma\sigma^\prime$ & $\sigma\pi^\prime$  & $\sigma\sigma^\prime$ & $\sigma\pi^\prime$ \\ 
\hline
\multirow{1}{*}{E1-E1} & $F^{1}$ & 0 & $1 \times 10^{-3}$ & 0 & $3 \times 10^{-5}$ & 0 & $1 \times 10^{-5}$ \\
\hline
\multirow{2}{*}{E2-E2}& $F^{1}$ & $2 \times 10^{-6}$ & $2 \times 10^{-6}$ & $2 \times 10^{-6}$ & $2 \times 10^{-6}$ & $2 \times 10^{-6}$ & $2 \times 10^{-6}$ \\ 
 & $F^{3}$ & $2 \times 10^{-6}$ & $2 \times 10^{-6}$ & $2 \times 10^{-6}$ & $2 \times 10^{-6}$ & $2 \times 10^{-6}$ & $2 \times 10^{-6}$ \\
\end{tabular}
\end{ruledtabular}
\caption{The results of $FDMNES$ calculations of XRS 
multipole intensities at the Mn $K$ edge, in units of 
electrons$^{2}$, in the cycloidal phase. All calculations 
were undertaken for azimuth $\psi=0$. The $E1-E2$ multipoles 
are negligible in the calculations because the Mn ions 
are located at cites with centre inversion symmetry in the $Pbnm$ space group.
The even order $E1-E1$ and $E2-E2$ and the time-even $E1-E2$ terms are explicitly zero
because they cannot arise from our magnetic structure. In the table zero means the 
calculated intensities were less than $1 \times 10^{-6}$ i.e. very small, not zero}
\label{tab:mn_cyc}
\end{center}
\end{table*}

\begin{table*}
\begin{center}
\begin{ruledtabular}
\begin{tabular}{|c|c|c|c|c|c|c|c|}
\multicolumn{8}{|c|}{Cycloidal Phase Tb $L_{3}$ edge}\\
\hline
Transition &  Multipole &  $(0,4\pm\mathrm q,1)$ &  $(0,4\pm\mathrm q,1)$ &  $(0,4\pm\mathrm q,0)$  & $(0,4\pm\mathrm q,0)$ &  $(0,3\pm\mathrm q,0)$ & $(0,3\pm\mathrm q,0)$ \\
& &  $\sigma\sigma^\prime$ & $\sigma\pi^\prime$  &  $\sigma\sigma^\prime$ & $\sigma\pi^\prime$  & $\sigma\sigma^\prime$ & $\sigma\pi^\prime$ \\ 
\hline
\multirow{1}{*}{E1-E1} & $F^{1}$ & 0 & $5 \times 10^{-3}$ & 0 & $1\times 10^{-3}$ & $1.4 \times 10^{-5}$ & $3.9 \times 10^{-4}$ \\
\hline
\multirow{2}{*}{E2-E2} & $F^{1}$ & 0 & 0 & 0 & $1.4 \times 10^{-6}$ & 0 & 0 \\
 & $F^{3}$ & 0 & 0 & 0 & 0 & 0 & 0 \\
 \hline
 \multirow{3}{*}{E1-E2} & $F^{1-}$ & 0 & 0 & 0 & $3 \times 10^{-6}$ & $5 \times 10^{-5}$ & 0 \\
 & $F^{2-}$ & 0 & 0 & 0 & 0 & $4 \times 10^{-6}$ & 0 \\
 & $F^{3-}$ & 0 & 0 & 0 & 0 & $6 \times 10^{-6}$ & 0 \\
\end{tabular}
\end{ruledtabular}
\caption{The results of $FDMNES$ calculations of all XRS multipole 
intensities at the Tb $L_{3}$ edge, in units of electrons$^{2}$, 
in the cycloidal phase. The $\pm$ for the $E1-E2$ multipoles refer to time even and time odd 
multipoles, respectively. All calculations were undertaken for azimuth $\psi=0$, 
except the $(0,3\pm\mathrm q,0)$ $\sigma\pi^\prime$ which was calculated 
at azimuth $\psi=90^\circ$. The even order $E1E1$ and $E2E2$ and the time-even $E1-E2$ terms are explicitly zero
because they cannot arise from our magnetic structure. In the table zero means the 
calculated intensities were less than $1 \times 10^{-6}$ i.e. very small, not zero}
\label{tab:tb_cyc}
\end{center}
\end{table*}

\clearpage
\newpage

\bibliography{tbmno3,homno3,rmn2o5,xrms,xrs}

\end{document}